\documentclass[onecolumn,showpacs,amsmath,amssymb,nofootinbib]
{revtex4}
\usepackage{amsmath}

\usepackage{physics}
\usepackage{subfigure}
\usepackage[T1]{fontenc}
\usepackage[utf8]{inputenc}
\usepackage{graphicx,dcolumn,bm,color,latexsym,amssymb}
\usepackage{import}
\usepackage{xifthen}
\usepackage{pdfpages}
\usepackage{transparent}
\usepackage{verbatim}

\definecolor{Blue}{rgb}{0.0,0.0,1}
\definecolor{Red}{rgb}{1,0.0,0.0}
\definecolor{Green}{rgb}{0,0.5,0.0}

\begin{document}
\title{The Heat Distribution of the Underdamped Langevin Equation.}


\author{Pedro V. Paraguass\'{u}}

\address{Departamento de F\'{i}sica, Pontif\'{i}cia Universidade Cat\'{o}lica\\ 22452-970, Rio de Janeiro, Brazil}

\author{Rui Aquino}
\address{Departamento de F\'{i}sica Te\'{o}rica, Universidade do Estado do Rio de Janeiro,\\ 20550-013, Rio de Janeiro, Brazil}

\author{Welles A.~M. Morgado}
\address{Departamento de F\'{i}sica, Pontif\'{i}cia Universidade Cat\'{o}lica\\ 22452-970, Rio de Janeiro, Brazil\\ and National  Institute of Science and Technology for Complex Systems, Brazil}

\date{January 2022}

\begin{abstract}
In Stochastic Thermodynamics, heat is a random variable with a probability distribution associated. Studies of the distribution of heat are mostly in the overdamped regime and in one dimension. Here we solve the heat distribution in the underdamped regime for three different cases: the free particle, the linear potential, and the harmonic potential. All of them in arbitrary dimensions. The results are exact and generalize known results in the literature.
\end{abstract}

\maketitle

\section{Introduction}
Stochastic Thermodynamics \cite{oliveira2020classical,ciliberto_experiments_2017,sekimoto2010stochastic,ryabov2015stochastic,seifert_stochastic_2019,peliti2021stochastic} is a recently developed  field that connects thermodynamics with the fluctuation world of small  out-of-equilibrium systems. Over the past two decades, many results were made available due to the discovery of fluctuation theorems \cite{seifert2005entropy,chernyak2006path,jarzynski2011equalities,gong2015jarzynski,seifert2012stochastic}, which are constraints on the nonequilibrium distribution of thermodynamic quantities like work, heat, and entropy.

Besides the general discoveries of fluctuation theorems, a more fundamental approach is to characterize a system in a nonequilibrium regime, which could need more effort than just finding constraints for the probabilities. Heat and other thermodynamic functionals are random variables and have associated probability distributions. The calculation of the probability distribution of a thermodynamic functional is in itself a difficult task, and only a few analytical results are possible. Exact results are found in the overdamped regime \cite{paraguassu_heat_2021,gupta_heat_2021,fogedby_heat_2020,goswami_heat_2019,crisanti_heat_2017,ghosal_distribution_2016,rosinberg_heat_2016,kim_heat_2014,kusmierz_heat_2014,saha_work_2014,chatterjee_single-molecule_2011,chatterjee_exact_2010,fogedby_heat_2009,imparato_probability_2008,imparato_work_2007,joubaud_fluctuation_2007}, where the inertia can be neglected due to the strong friction.    For the underdamped limit where the inertia is important, to the best of our knowledge,  the list is shorter \cite{Kwon2013,taniguchi2008inertial,sabhapandit_heat_2012,PhysRevE.90.052116,munakata_entropy_2012,rosinberg_stochastic_2017,crisanti_heat_2017}, and is restricted for one dimensional systems. Interestingly, the most common method used is the path integral technique of a stochastic process \cite{wio2013path,cugliandolo2017rules,cugliandolo2019building}.

  The limited number of studies in the underdamped limit can be justified by the simplicity of the path integral formalism of the overdamped regime, and also by the fact that many systems of interest are quite well represented by overdamped models. Moreover, the one-dimensional approximation simplifies the calculations without losing the fundamental aspects of physics. When dealing with the underdamped limit, one of the main difficulties that arise is the appearance of second derivatives in the action \cite{kleinert_path_1986,chouchaoui_path_1993,suassuna_path_2021,rosinberg_stochastic_2017,munakata_entropy_2012}, breaking the correspondence with well-known quantum mechanical results \cite{chaichian2018path}. Moreover, besides the one dimension case being a good approximation for the physics of the Brownian particle, some features only appear in higher dimensions. Herein, we shall work out a way to deal with these difficulties, working with the underdamped case in arbitrary dimensions. Beyond that, the overdamped regime is a good approximation because experimentally it describes the behavior of a colloidal particle in a harmonic trap \cite{douarche_work_2006}. However, there are experimental cases where inertia cannot be neglected. The Brownian Carnot machine \cite{martinez2016brownian, holubec2021fluctuations} is an example of such a situation.

In this work, we draw attention to the most fundamental consequences of the underdamped Langevin equation, we investigate the heat distribution of the underdamped particle in $n-$dimension case. Our results extend the list of exact results for the heat in the underdamped regime, comparing with known results from the literature \cite{rosinberg_stochastic_2017,chatterjee_exact_2010,munakata_entropy_2012}. We investigate three cases; firstly, we revisit a well-studied system, which is the free particle in contact with a heat bath in the underdamped regime \cite{lemons1997paul,sekimoto1998langevin}; secondly, we include a constant linear potential in the system, which can be interpreted as an external force \cite{singh_onsager-machlup_2008,ciliberto_experiments_2017}; thirdly, we study the harmonic trap, which is widespread in the literature since optical tweezers system can be modeled that way \cite{douarche_work_2006,ciliberto_experiments_2017}. The investigation is carried by the calculation of the characteristic function, the central moments, skewness and kurtosis \cite{wampler2021skewness}, to finally derive the heat distribution.
For these potentials, the transitional probability can be calculated analytically, and we review its calculation via path integrals \cite{wio2013path, suassuna_path_2021}. Despite the simplicity of these systems, finding the heat distribution is quite nontrivial, and analytical results for the heat distribution are only possible for the first and third case. However, the characteristic function can be obtained analytically for arbitrary dimensions, and all heat distributions here are exact, in the sense that we can use numerical integration to find the heat distribution. We compared our results with results from the literature, finding similarities between underdamped and overdamped systems. Moreover, the distributions for cases with potentials are consistent with the free case as we turn off interactions.

One important difference for the heat flow, between the overdamped and the underdamped cases, is the appearance of the kinetic energy contribution in the last case. This contribution is often called heat leakage \cite{arold_heat_2018}, since it is not captured in the overdamped case.  It has been shown in the literature, that in some cases, while the dynamics are overdamped, the energetics is not. In these events, the energetics of the overdamped regime seems not to be the limiting case for the underdamped model. \cite{celani_anomalous_2012,dechant_underdamped_2017, bo2013entropic, schmiedl2007efficiency, plata2020building}. Herein, we highlight that the heat distribution of the free particle is one of these events.

Furthermore, the distributions derived are in agreement with the work of Rosinberg et al. \cite{rosinberg_heat_2016}, where it has been shown that the heat distribution in the underdamped regime obeys an integral fluctuation theorem, that can give estimates for the exponential tail of such distributions. The distributions found here demonstrate this exponential tail. Also, recently, Fodgeby has shown that the heat distribution in the underdamped regime of a harmonic trap has a symmetric exponential behavior in the equilibrium limit \cite{Fogedby_2020}. We recover his result in the asymptotic limit in our third case and extend it to higher dimensions.

This paper is organized as follows. In Sec.~\ref{sec2} we  review  the stochastic ther\-mo\-dy\-na\-mics of the Brownian particle. In Sec.~\ref{sec3} we calculate the characteristic function for the free, linear, and harmonic case as a preliminary step to obtain the heat distribution. We also study the central moments of the heat in these cases. In Sec.~\ref{sec4} we derive the heat distribution for the studied cases. Moreover, we investigate the role of the isotropy of the space in the linear case. In \ref{sec5} we discuss the results and compare qualitatively the distribution.  We finish in Sec.~\ref{sec6} with a conclusion of the results.

\section{Stochastic Thermodynamics of Brownian Motion}
\label{sec2}
The Brownian particle is one of the most standard systems in the domain of Stochastic Thermodynamics. Before deriving the heat distribution for 3 different cases, we give a brief review of the thermodynamics of such systems. Written  in vector form, the definitions given here are valid for general dimensions.

The Brownian particle, which is in contact with a heat bath, evolves in time according to the generalized Langevin Equation \cite{coffey2012langevin}
\begin{equation}
    m\frac{d^2\vec{r}}{dt^2}= -\gamma \frac{d\vec{r}}{dt} -\nabla V(\vec{r},t) + \vec{\eta}(t).\label{genLa}
\end{equation}
This corresponds to Newton's second law of classical mechanics with a stochastic nature, due to the bath force $\eta(t)$. This force models the interaction between the particle and the fast degrees of freedom of the bath~\cite{Mori1965}. For a thermal bath, in the paradigmatic case of Stochastic Thermodynamics, this force is modeled by a Gaussian white noise. Then, the bath generated force has the statistical properties (all other cumulants are null):
\begin{equation}
    \langle \eta_i(t)\eta_j(t') \rangle = 2\gamma T \delta_{ij}\delta(t-t'),\;\;\;\; \langle \vec{\eta}(t)\rangle = \vec 0,
\end{equation}
where $T$ is the temperature of the heat bath, and $\gamma$ is the frictional coupling constant. Note that we are working here with $k_B=1$. 
The other two forces on the right side in \ref{genLa} are the frictional force $-\gamma \dot{\vec{r}}$ which comes from the slow degrees of freedom of the bath, and the force $-\nabla V(\vec{r},t)$ which is a deterministic force derived from an internal potential $V(\vec{r},t)$. Here we are considering only deterministic forces which come from an internal potential. However, there could exist external protocols.  The time dependence of the conservative internal potential allows for the inclusion of an external protocol in the parameters of the potential. As an example, the deterministic force can be an electric field generated by a capacitor-like device, with time variation of the voltage \cite{singh_onsager-machlup_2008,Rossnagel2016}, or be a harmonic potential where the spring constant might vary in time according to  some protocol \cite{douarche_work_2006,martinez2016brownian}. These are similar examples of the systems that will be studied in the next section.

For the  Stochastic Thermodynamics of the above system, we start defining the heat exchanged between the particle and the bath. Following Sekimoto\cite{sekimoto2010stochastic} we have
\begin{equation}
    Q[\vec{r}]= \int_0^\tau \left(-\gamma \frac{d\vec{r}}{dt} + \vec{\eta}(t)\right)\cdot \frac{d\vec{r}}{dt} dt, 
\end{equation}
where the first term takes into account the energy lost by the particle to the reservoir, while the second term is the energy acquired by the particle from the same reservoir. The functional dependence on the stochastic variable $\vec{r}(t)$ gives a stochastic nature to the heat. The stochastic nature of the thermodynamic variables, such as heat, work, and entropy, is what justifies the name Stochastic Thermodynamics \cite{ryabov2015stochastic}.  The above formula represents  the heat exchange during a time interval of $t\in[0,\tau]$. Using the Langevin Eq.~(\ref{genLa}) we have 
\begin{equation}
    Q[\vec{r}]=\int_0^\tau \left(m\frac{d^2\vec{r}}{dt^2} +\nabla V(\vec{r},t) \right)\cdot \frac{d\vec{r}}{dt} dt = \frac{1}{2}m(v_\tau^2-v_0^2)+ \int_0^\tau \nabla V(\vec{r},t)\cdot\frac{d\vec{r}}{dt}dt. \label{1law}
\end{equation}
Notice that $Q[\vec{r}(\tau)]$ depends on the variation of the kinetic energy, where we define $\dot{ \vec{r}}(\tau)=\vec{v}(\tau)=v_\tau$, $\dot {\vec{r}}(0)=\vec{v}_0$. Herein, we only deal with cases where the conservative potential is time independent $V(\vec{r},t)\rightarrow V(\vec{r})$. In this case, the heat becomes
\begin{equation}
    Q[\vec{r}]= \frac{1}{2}m(v_\tau^2-v_0^2) {+} \Delta V  = \Delta E \label{equacao6}
\end{equation}
where in the second equality we define, \[\Delta E = \frac{1}{2}m(v_\tau^2-v_0^2)  {+} \Delta V , \] as the variation of the internal energy of the particle, and $\Delta V = V(\vec{r}_\tau)-V(\vec{r}_0)$. Note that Eq.~(\ref{equacao6}) is the first law of thermodynamics. Herein, we shall concentrate in characterizing the heat, which can be expressed as changes of the internal energy, as expressed in the First Law.  The fact the the boundary terms are also random variables yields  a non-trivial contribution for the probabilities distributions.

\section{Characteristic function in $n-$dim and central moments}\label{sec3}

The characteristic function is the Fourier transform of the heat distribution and gives us all the moments of the distribution. Thus, before the calculation of the heat distribution is instructive to solve for the characteristic function. Moreover, the characteristic function can be calculated analytically for the heat in the underdamped system only for the three different cases: harmonic, free particle, and linear potential. This is possible because the transitional probability in these cases can be solved analytically via Fokker-Planck or path integrals. Here, we solved via path integrals \cite{wio2013path,chaichian2018path,suassuna_path_2021}, and the calculations are given in the appendix. 

Since the heat only depends on the boundary terms, the characteristic function in $n$ dimensions will be given by
\begin{equation}
    Z(\lambda) = \langle e^{-i\lambda Q[{r}]}\rangle^n,\label{zlambs}
\end{equation}
where $n$ is the dimension of the model that we are working on, and the average is over the degrees of the position and velocities degrees of freedom. (the calculation via path integral is given in the appendix \ref{apA}). This is only possible if (i) the noise is uncorrelated, meaning that the positions and velocities in any direction are also uncorrelated (ii) the potential is a sum of each component contribution having the same constants, and thus the Langevin equation for each component is independent.

Let us now calculate the moments of the distribution for the harmonic, free, and linear cases. Which can be calculated just by
\begin{equation}
    i^n  \frac{\partial^n Z(\lambda)}{\partial \lambda^             n}\bigg|_{\lambda=0}=\langle Q^n \rangle.
\end{equation}
Moreover, since the characteristic function is just the multiplication of the one-dimensional case, the dimension will only affect the results as a multiplicative constant.

\subsection{Harmonic and free particle case}

The free particle case can be obtained by turning off the potential of the harmonic case (or the linear). Here we will start with the harmonic case, with the potential $V(r)=\frac{kr^2}{2}$, where $r^2=|\vec r|^2$, the heat will be
\begin{equation}
    Q[\vec r] = \frac{1}{2}m\left(v_\tau^2-v_0^2\right) + \frac{1}{2}k \left(r_\tau^2-r_0^2\right),
\end{equation}
thus, assuming an equilibrium initial distribution for the velocities and positions, since the potential is time independent and experimentally its a state that the particle will achieve, the formula in Eq.~(\ref{zlambs}) will be just Gaussian integrals, (see appendix \ref{apB}) and the characteristic function will be
\begin{equation}
   Z_{h}(\lambda) = \alpha ^{n/2} e^{\frac{\gamma  n \tau }{2 m}} \left(\alpha  e^{\frac{\gamma  \tau }{m}}+2 \lambda ^2 T^2 \left(\alpha  \sinh \left(\frac{\gamma  \tau }{m}\right)+\lambda ^2 T^2+1\right)\right)^{-\frac{n}{2}},\label{zharm}
\end{equation}
where $\alpha$ is a constant that depends on the parameters of the problem, 
\begin{equation}
\alpha = \frac{\left(4 k m-\gamma ^2\right) }{ \left(\gamma ^2 \cosh \left(\frac{\tau  \sqrt{\gamma ^2-4 k m}}{m}\right)+\left(4 k m-\gamma ^2\right) \cosh \left(\frac{\gamma  \tau }{m}\right)-4 k m\right)}.
\end{equation}
The derivation of the characteristic function of the harmonic case is given in appendix~\ref{apB}.

\begin{figure}
    \centering
    \includegraphics[width = 15cm]{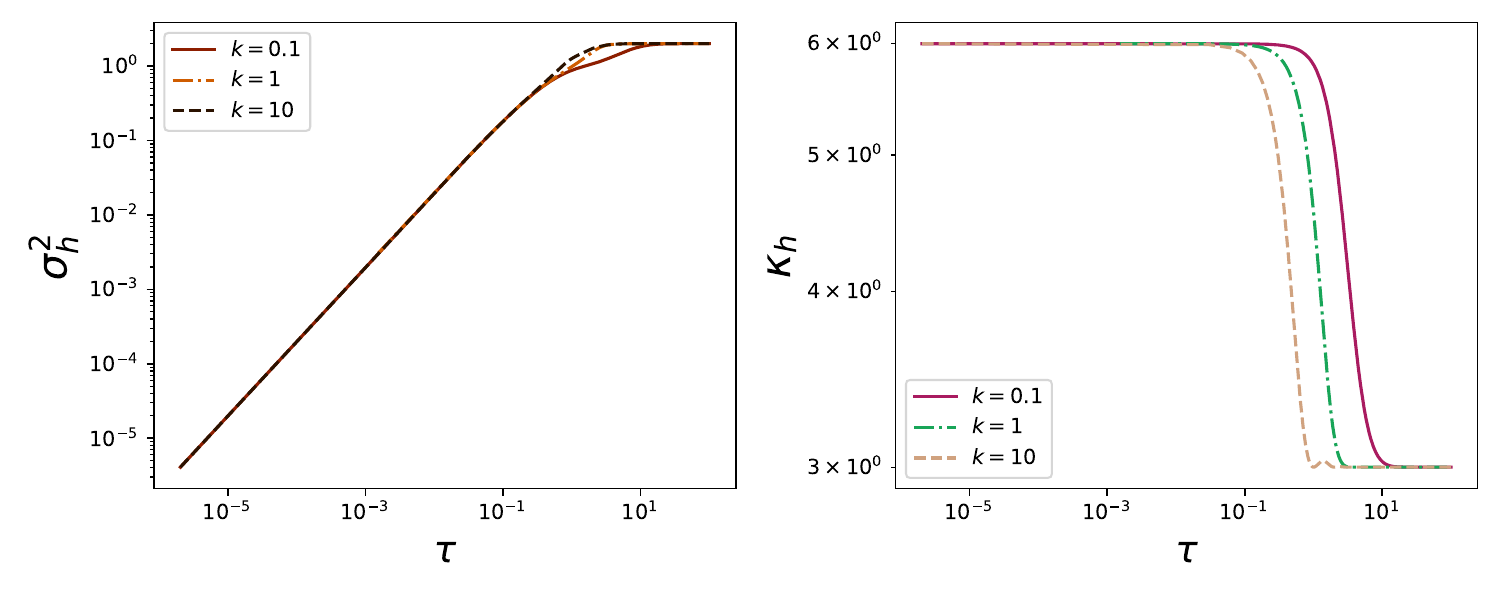}
    \caption{Left: Variance of the harmonic case over time for different values of $k$. All the remaining parameters are set to one. Right) Kurtosis of the harmonic case over time for different values of $k$.}
    \label{varkurharm}
\end{figure}

By calculating the moments, we can obtain the central moments, the mean, variance, skewness and excess kurtosis. The mean $\mu=\langle Q\rangle$ and skewness $\mu_3= \langle\left(Q-\langle Q\rangle \right)^3\rangle$ are null. Meaning that the distribution will be symmetric around the mean, as expected, since the potential is symmetric and thus the heat flux can not have a preferred direction. The variance $\sigma^2=\langle \left(Q-\langle Q\rangle \right)^2 \rangle$ and excess kurtosis $\kappa ={\langle \left(Q-\langle Q \rangle\right)^4\rangle}\sigma^{-4}-3$ are
\begin{equation}
    \sigma^2_h= \alpha^{-1}{2 n T^2 e^{-\frac{\gamma  \tau }{m}} \left(\alpha  \sinh \left(\frac{\gamma  \tau }{m}\right)+1\right)},
\end{equation}
\begin{equation}
    \kappa_h = n^{-1}{6 \left(1-\frac{\alpha  e^{\frac{\gamma  \tau }{m}}}{\left(\alpha  \sinh \left(\frac{\gamma  \tau }{m}\right)+1\right)^2}\right)}.\label{kurtfree}
\end{equation}
The variance is positive and increases in time, thus the distribution becomes broader as time pass. Moreover, the variance increases as we increase the spatial dimension. While the kurtosis is compacted by the dimension, that is, the kurtosis decreases as the dimension increases. Nevertheless, the term inside the parenthesis of eq.~(\ref{kurtfree}) kurtosis is always positive and greater than 3, as can be checked by hand for a different set of parameters, since $k>0,m>0,\gamma>0$. And, as long $n$ is a positive integer, the excess kurtosis is always positive, meaning that the heat distribution in the harmonic case is Leptokurtic. Implying that the distribution has fatter tails, having more probabilities to occur rare events than a normal distribution. We plot the variance and kurtosis respectively in the left and right of figure~\ref{varkurharm}, for $n=1$ since $n$ only affects as a multiplicative constant. One can see that the behavior is qualitative the same for different values of $k$. The only difference is, as we increase $k$ the averages become stationary faster, but the asymptotic values are the same.

To obtain the free particle case, we can take $k\rightarrow 0$. The mean and skewness are also zero, while the characteristic function is now
\begin{equation}
    Z_{ f}(\lambda) = \left(\frac{2\lambda^2T^2}{\coth{\frac{\gamma \tau}{m}}+1}+1\right)^{-\frac{n}{2}}.\label{zfree}
\end{equation}
The variance and kurtosis can be obtained by the limit $k\rightarrow 0$ or by taking the derivatives of Eq.~(\ref{zfree}) and are given by
\begin{equation}
    \sigma^2_f=\frac{2 n T^2}{\coth \left(\frac{\gamma  \tau }{m}\right)+1},
\end{equation}
\begin{equation}
    \kappa_f = \frac{6}{n}.
\end{equation}
The variance is always positive, and increasing in time and has the same qualitative behavior of the harmonic case. Interesting, the kurtosis is constant in time, and is also positive, and thus Leptokurtic.

\subsection{Linear case}
The linear case can be obtained by a Galilean transformation of the velocity on the free particle case. However, we will calculate separately the transitional probability that depends on the velocity and position (see appendix \ref{apB}). The linear potential is $V(\vec{r})=-\vec\kappa\cdot \vec r$, and the heat is now
\begin{equation}
    Q[\vec r] = \Delta K -\vec {\kappa}\cdot\left(\vec r_\tau-\vec r_0\right),
\end{equation}
where $\vec{\kappa}$ is a $n$ dimensional vector, in which all the components are the constant of the potential $\kappa$. 

The characteristic function is
\begin{equation}
   Z_l(\lambda)= \frac{\exp \left(\frac{n\gamma  \tau }{2 m}+nf_2(\lambda,\kappa)\right)}{\sqrt{\cosh \left(\frac{\gamma  \tau }{m}\right)+\left(2 \lambda ^2 T^2+1\right) \sinh \left(\frac{\gamma  \tau }{m}\right)}^n},\label{zlin}
\end{equation}
where $f_2(\lambda,\kappa)$ is given in the appendix \ref{apC}. Here, we do not need to assume any initial distribution, since by integrating first the final boundary conditions, the dependence on the initials are removed. This occurs due the transnational symmetry of the linear case. Note that, due to $f_2(\lambda,\kappa)$ the mean and skewness are no longer null. Physically, is due to the asymmetry of the linear potential. 

The mean is
\begin{equation}
    \mu_l = \frac{\kappa ^2 n }{2 \gamma ^2}\left(m \left(e^{-\frac{2 \gamma  \tau }{m}}-4 e^{-\frac{\gamma  \tau }{m}}+3\right)-2 \gamma  \tau \right),
\end{equation}
which decreases as time pass, and its asymptotic time limit gives $\mu_l\rightarrow-\infty$. Note that the mean is independent of the signal of $\kappa$, as should be, since the heat can not depend on the direction of the force applied. 
The variance has a more complicated formula and is
\begin{equation}
    \sigma_l^2=\frac{n }{\gamma ^2}T e^{-\frac{2 \gamma  \tau }{m}} \left(\kappa ^2 \left(4 m e^{\frac{\gamma  \tau }{m}}+e^{\frac{2 \gamma  \tau }{m}} (2 \gamma  \tau -3 m)-m\right)+\gamma ^2 T \left(e^{\frac{2 \gamma  \tau }{m}}-1\right)\right).
\end{equation}
Note that, for $\tau\rightarrow\infty$ the variance diverges. This divergence, together with the divergence of the mean, means that the heat distribution is flat for an asymptotic time, due to the linear potential. Moreover, note that the mean and variance only depend on $\kappa^2$, meaning that statistical behavior of the heat does not depend on the signal of the force, as it should be. In the next section, we will further discuss the direction and signal of $\kappa$ in the heat distribution.

\begin{figure}
    \centering
    \includegraphics[width=15cm]{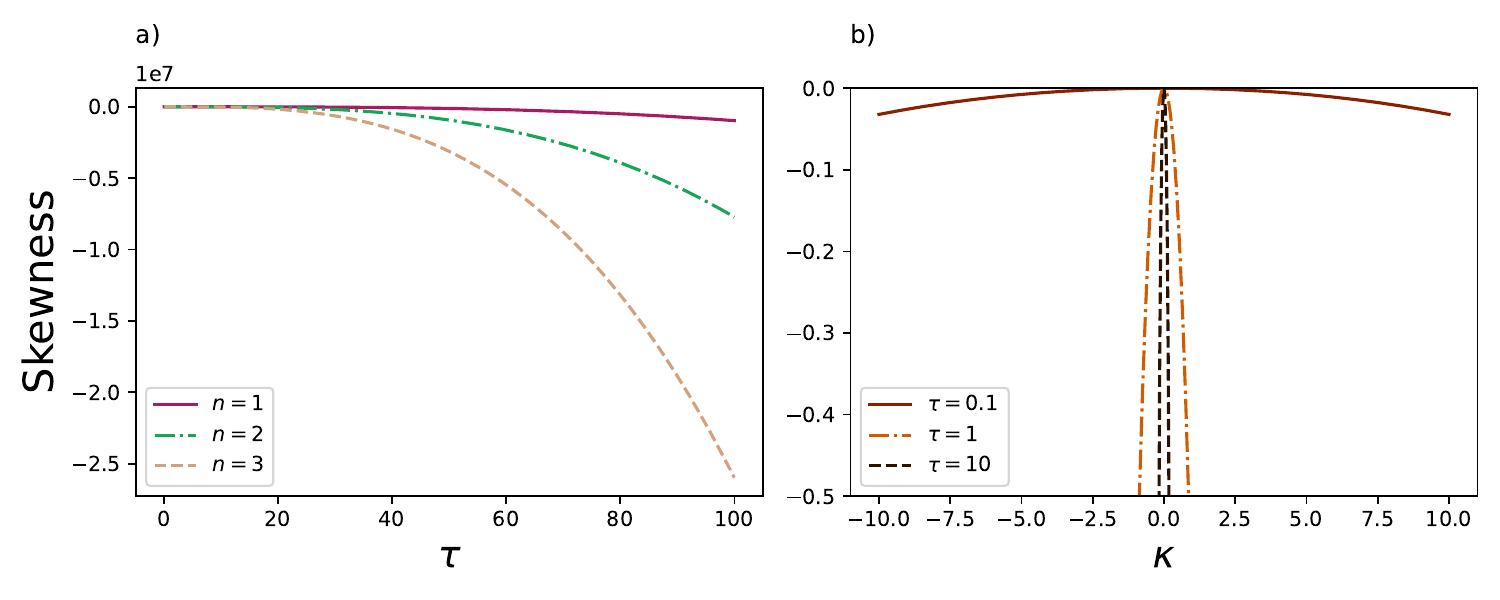}
    \caption{a) Skewness for the linear case over time for different dimensions. All the remaining parameters are set to one. b) Skewness over $\kappa$ for different times. All the remaining parameters are set to one. Note that the skewness is independet of the signal of $\kappa$.}
    \label{skewnes}
\end{figure}

The skewness can be obtained analytically and has a long formula, thus we decide to omit it in the presentation. The skewness tells us about the asymmetry of the heat distribution. In figure~\ref{skewnes} a) one can see that for the three dimensions, the skewness always starts in zero (meaning that the heat distribution is symmetric) and become negative as time pass, according to the asymptotic behavior of the mean. Moreover, for higher dimensions, the skewness decreases more rapidly in time. By changing the force constant $\kappa$, one can see in figure~\ref{skewnes} b) that the skewness is also independent of the signal of $\kappa$. And decreases as we increase the modulus of $\kappa$.

The excess of kurtosis has a more simply formula and is given by
\begin{equation}
   \kappa_l = \frac{48 \gamma ^2 T e^{\frac{2 \gamma  \tau }{m}} \sinh ^2\left(\frac{\gamma  \tau }{2 m}\right) \left(-2 \kappa ^2 m+\cosh \left(\frac{\gamma  \tau }{m}\right) \left(2 \kappa ^2 m+\gamma ^2 T\right)+\gamma ^2 T\right)}{n \left(-4 \kappa ^2 m e^{\frac{\gamma  \tau }{m}}+\kappa ^2 m-e^{\frac{2 \gamma  \tau }{m}} \left(\kappa ^2 (2 \gamma  \tau -3 m)+\gamma ^2 T\right)+\gamma ^2 T\right)^2}.
\end{equation}
Note that, the kurtosis of the linear potential also decreases as we increase the dimensions. Moreover, like the skewness, variance, and mean, the kurtosis, $\kappa_l$, is independent of the signal of the force constant $\kappa$. Furthermore, the kurtosis is always positive, meaning that the distribution is also Leptokurtic.

\section{Heat Distributions}\label{sec4}
In general, to obtain the heat distribution, the starting point is the conditional heat distribution, which is
\begin{equation}
    P(Q|Q=Q[\vec r]) = \delta(Q-Q[{\vec r}])
\end{equation}
ensuring the fact that the random variable $Q$ is given by the functional formula $Q[\vec{r}]$. Integrating out the conditional, that is, $\langle \delta(Q-Q[\vec r]) \rangle$, we have the desired heat distribution
\begin{equation}
    P(Q)= \int d^3\vec{v}_\tau d^3\vec{r}_\tau \int d^3\vec{v}_0 d^3\vec{r}_0 P_0(\vec{r}_0,\vec{v_0}) \int \mathcal{D}\vec{r} e^{-\frac{1}{4\gamma T}\mathcal{A}[\vec{r}]}\delta(Q-Q[\vec{r}]).\label{heatpath}
\end{equation}
The Dirac delta can be rewritten as an integral, giving
\begin{equation}
    P(Q)= \int \frac{d\lambda}{2\pi} e^{i\lambda Q} Z(\lambda),\label{fourie}
\end{equation}
where $Z(\lambda)$ was already given in eq.~(\ref{zlambs}). Also, the above formula justifies the formula in eq.~(\ref{zlambs}), where the average is the integration over all random variables as we see in eq.~(\ref{heatpath}).

Let us now analyze the three paradigmatic cases for arbitrary dimensions.

\subsection{Free particle}

For the free particle, due eq.~\ref{fourie}, the Fourier transform of the characteristic function, eq.~\ref{zfree}, can be obtained analytically for any dimension. Thus, the heat distribution for $n-$dim are
\begin{equation}
    P_f(Q) = \frac{2^{\frac{1}{4}-\frac{3 n}{4}} T^{\frac{1}{2} (-n-1)}}{\sqrt{\pi } \Gamma \left(\frac{n}{2}\right)} | Q| ^{\frac{n-1}{2}} \left(\coth \left(\frac{\gamma  \tau }{m}\right)+1\right)^{\frac{n+1}{4}} K_{\frac{n-1}{2}}\left(\frac{| Q|  \sqrt{\coth \left(\frac{\gamma  \tau }{m}\right)+1}}{\sqrt{2} T}\right).
\end{equation}
The heat distribution for the free particle is a Besselian probability distribution. Note that only for $n=1$ we no longer have the modulus of $Q$ outside the Bessel function, meaning that for $Q=0$ the distribution is divergent. 

For one dimension we have
\begin{equation}
    P_f(Q)^{1D}=\frac{\sqrt{\coth \left(\frac{\gamma  \tau }{m}\right)+1}}{\pi \sqrt{2}T} \;K_0\left(\frac{| Q|  \sqrt{\coth \left(\frac{\gamma  \tau }{m}\right)+1}}{\sqrt{2} T}\right).\label{heatdist2}
\end{equation}
The result Eq.~(\ref{heatdist2}) is the heat distribution for an underdamped free particle in contact with a heat bath of temperature $T$. This distribution is mathematically equivalent to the heat distribution of an overdamped particle diffusing in a harmonic potential \cite{chatterjee_exact_2010}, this means, for example, that a pollen grain and an acrylic particle under optical tweezers, absorb heat with the same statistics, albeit for distinctly defined  constant. Due to the modified Bessel function, we have an exponential tail in the distribution, as expected \cite{rosinberg_heat_2016}.

Other interesting cases are the two and three-dimensional cases, where the distributions are
\begin{eqnarray}
    P_f(Q)^{2D} = \frac{ e^{-\frac{| Q|  \sqrt{\coth \left(\frac{\gamma  \tau }{m}\right)+1}}{\sqrt{2} T}}\,\sqrt{\coth \left(\frac{\gamma  \tau }{m}\right)+1}}{2 \sqrt{2} T},\\
    P_f(Q)^{3D}=\frac{\left(\coth \left(\frac{\gamma  \tau }{m}\right)+1\right)}{2\pi T^2} |Q| K_1\left(\frac{|Q|\sqrt{\coth \left(\frac{\gamma  \tau }{m}\right)+1}}{\sqrt{2}T}\right).
\end{eqnarray}
The two-dimensional case reduces to a Laplacian distribution, while the three-dimensional case still has a Bessel function, but now of first order.  

For the free particle, in all dimensions, we have a symmetric distribution in the argument $Q$, this means that the particle does not tend to gain or to lose energy from the heat bath. For 1D, it is interesting to note that this case has a mathematical equivalence with the overdamped particle in a quadratic potential \cite{chatterjee_exact_2010}, and this equivalence comes from the fact that we can write the second-order stochastic differential equation for $x(t)$ as a first-order stochastic differential equation in $v(t)$. Thus, our findings in 3D and 2D serve as the heat distribution for the overdamped harmonic particle. For the 2D and 1D cases, there is a peak in $Q=0$ which comes from the Bessel function K and the exponential, while in 3D this peak is smoothed by the modulus of $Q$ outside the Bessel function.

The asymptotic time limit $\tau\rightarrow\infty$ and the overdamped limit $\gamma/m \rightarrow \infty$ are equivalently the same, and the distribution is
\begin{equation}
  \lim_{\gamma/m\rightarrow\infty} P_f(Q) =\lim_{\tau\rightarrow\infty} P_f(Q) = \frac{2^{\frac{1}{2}-\frac{n}{2}} T^{\frac{1}{2} (-n-1)} }{\sqrt{\pi } \Gamma \left(\frac{n}{2}\right)}| Q| ^{\frac{n-1}{2}} K_{\frac{n-1}{2}}\left(\frac{| Q| }{T}\right).
\end{equation}
Unexpectedly, the distributions we derived do not have a direct correspondence with the overdamped case. The results of the heat distribution are different if we take the overdamped limit before in the dynamics~\cite{NascimentoMorgado2019}. This non correspondence is in current study in the literature \cite{arold_heat_2018,celani_anomalous_2012,dechant_underdamped_2017,  bo2013entropic, schmiedl2007efficiency, plata2020building}. We pointed out, for the case discussed here, that there is a need for further investigations on this problem.

\subsection{Harmonic potential}

\begin{figure}
    \centering
    \includegraphics[width=17cm]{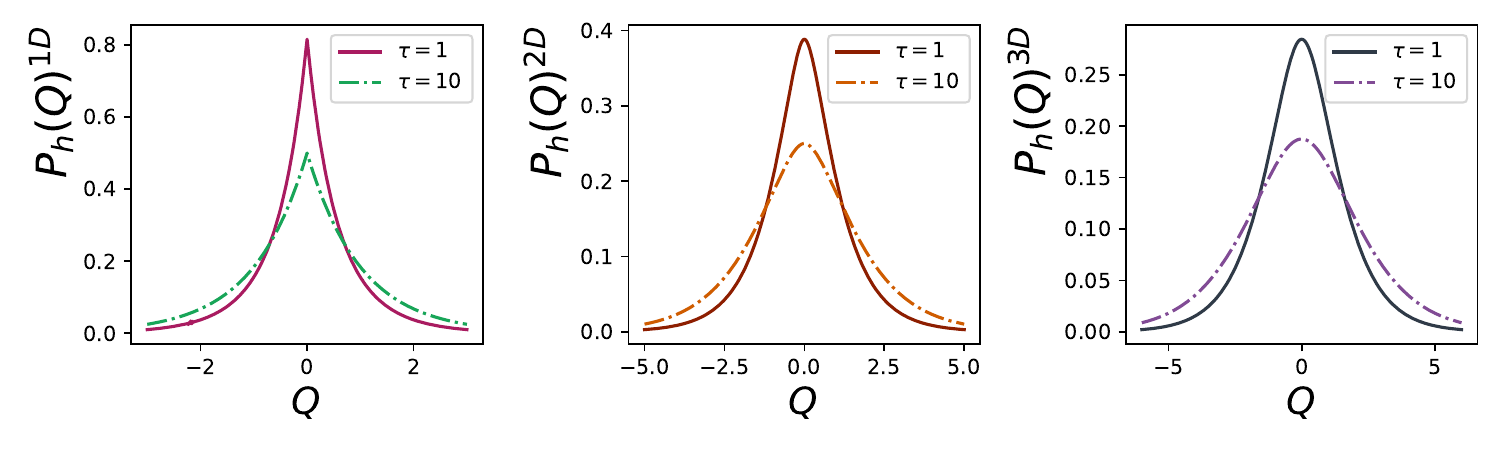}
    \caption{Heat distributions for the harmonic case in 1, 2, and 3 dimensions. All remaining parameters are set to one. Note the smoothness behavior of the distribution as we increase the dimension. The 2D potential is obtained analytically, while the 1D and 3D are solved by numerical integration of the characteristic function.}
    \label{harm}
\end{figure}

The characteristic function for the harmonic potential does not have an analytical solution since we have a fourth-order polynomial in the denominator inside the square root. However, for even dimensions, that is $n=2N$ with $N$ being a positive integer, the distribution reduces to the integral
\begin{equation}
    P_h(Q) = \int \frac{d\lambda}{2 \pi} e^{i\lambda Q} \alpha ^n e^{\frac{\gamma  n \tau }{m}} \left(\alpha  e^{\frac{\gamma  \tau }{m}}+2 \lambda ^2 T^2 \left(\alpha  \sinh \left(\frac{\gamma  \tau }{m}\right)+\lambda ^2 T^2+1\right)\right)^{-n},
\end{equation}
that can be calculated analytically by Residue theorem. To illustrate, for $N=1$, we are in 2 spatial dimensions and the integral becomes
\begin{equation}
    P_h(Q)^{2D}= \frac{\alpha T^2}{2km} e^{\frac{\gamma\tau}{m}} \int \frac{d\lambda}{2\pi}\frac{e^{i\lambda Q}}{\frac{1}{2} \alpha e^{\frac{\gamma  \tau }{m}}+\lambda ^2 T^2 \left(\alpha \sinh \left(\frac{\gamma  \tau }{m}\right)+1\right)+\lambda ^4 T^4}.
\end{equation}
The above equation can be solved analytically through the residue theorem. Calling $b=\alpha\sinh\left(\frac{\gamma\tau}{m}\right)+1$ and $c=\frac{1}{2}\alpha e^{\frac{\gamma \tau}{m}}$, it can be shown that $c>b$ and $b,c$ are positive real constants, and we make $\lambda \rightarrow T\lambda$. Therefore, we need to solve the integral over the complex plane
\begin{equation}
    \frac{1}{T}\int \frac{d\lambda}{2\pi} \frac{e^{i\lambda {Q}/T}}{\lambda^4+b\lambda^2+c}= \frac{1}{T} \prod_{i=1}^4\oint \frac{e^{i\lambda Q/T}}{(\lambda-\lambda_i)} \frac{d\lambda}{2\pi} = i \sum_j\,R_j
\end{equation}
where $R_j$ are the residues of the $\lambda_i$'s poles, that are
\begin{equation}
    \lambda_1 = -\frac{i}{\sqrt{2}}\sqrt{b+i\sqrt{4c-b^2},\;\;} \lambda_4= \frac{i}{\sqrt{2}}\sqrt{b-i\sqrt{4c-b^2}},\;\;\lambda_2=-\lambda_1,\;\;\lambda_3=-\lambda_2.\label{poles}
\end{equation}
Due to $b,c>0$ with $c>b$, one can see that $\lambda_1$ and $\lambda_4$ are in the upper half of the complex plane while $\lambda_2$ and $\lambda_3$ are in the lower half of the complex plane. Thus, by Residues' theorem, the integral becomes
\begin{equation}
  i \sum_j\,R_j =\frac{i}{2T(\lambda_4^2-\lambda_1^2)}  \left[\left( \frac{e^{i\lambda_4Q/T}}{\lambda_4}  -\frac{e^{i\lambda_1Q/T}}{\lambda_1} \right)\Theta(Q) + \left( \frac{e^{-i\lambda_1Q/T}}{\lambda_1}  -\frac{e^{-i\lambda_4Q/T}}{\lambda_4} \right)\Theta(-Q) \right].
\end{equation}
Therefore, the heat distribution is given by
\begin{equation}
    P_h(Q)^{2D}= \frac{\alpha T}{4km} \frac{ie^{\frac{\gamma\tau}{m}}}{(\lambda_4^2-\lambda_1^2)}\left[\left( \frac{e^{i\lambda_4Q/T}}{\lambda_4}  -\frac{e^{i\lambda_1Q/T}}{\lambda_1} \right)\Theta(Q) + \left( \frac{e^{-i\lambda_1Q/T}}{\lambda_1}  -\frac{e^{-i\lambda_4Q/T}}{\lambda_4} \right)\Theta(-Q) \right],
\end{equation}
with $\lambda_i$s given in eq.~(\ref{poles}).
Despite the imaginary number, the result is real as it can be seen in figure \ref{harm}. We also solve numerically for the 1D and 3D case and plot the results in figure~\ref{harm}. We notice that this distribution has a parallel behavior with the heat distribution of two overdamped particles in contact with two heat baths \cite{ghosal_distribution_2016}. This correspondence comes from the similar dependence of the exchanged heat in the four boundary terms of the respective two cases. Differently from the previous cases, the correspondence is not in the Langevin equation, it is in the similar functional form of the heat in both cases. As pointed out at the beginning of this section, it is a particular case of the models discussed in \cite{munakata_entropy_2012,rosinberg_stochastic_2017}. 

Nevertheless, for any dimension, the harmonic heat distribution can be solved analytically
\begin{equation}
   \lim_{\tau\rightarrow\infty} P_{h}(Q) = \frac{2^{\frac{1}{2}-n} T^{-n-\frac{1}{2}} }{\sqrt{\pi } \Gamma (n)}| Q| ^{n-\frac{1}{2}} K_{\frac{1}{2}-n}\left(\frac{| Q| }{T}\right),
\end{equation}
and is almost the same of the free case. The only difference are in the parameter $\frac{1}{2}-n$, while in the free case this parameter appear as $\frac{1}{2}(1-n)$. Therefore, for $n=2N$, the asymptotic harmonic distribution is equal to the asymptotic free distribution for $n=N$. Thus, the asymptotic $n$-dim harmonic case is equal to the asymptotic $2n-$dim free particle case. To illustrate, the asymptotic heat distribution for the harmonic in 2D and 3D are
\begin{equation}
    \lim_{\tau \rightarrow \infty}P_h(Q)^{1D} = \frac{1}{4\pi T^2}\left(T+|Q|\right)e^{-\frac{|Q|}{T}},
\end{equation}
\begin{equation}
    \lim_{\tau \rightarrow \infty}P_h(Q)^{3D} =  \frac{1}{16T}\left(\frac{Q^2}{T^2}+3+3\frac{|Q|}{T}\right)\exp\left(-\frac{|Q|}{T}\right).
\end{equation}

For the Langevin harmonic oscillator case, we found a distribution qualitatively similar to the free particle. Even though in odd dimensions we cannot write the distribution in terms of known functions, in general, the  Fourier transform of the characteristic function in eq.~(\ref{zharm}) has similar behavior the Fourier transform of eq.~(\ref{zfree}): both are symmetric in $\lambda$, with a polynomial dependence inside the square root of the denominator. The big difference is a fourth-order polynomial in the denominator of eq.~(\ref{zharm}), while in eq.~(\ref{zfree}) we have a second-order polynomial. This denominator can be solved in the even dimension case analytically, because the square root is removed, and the integral is simply solved by the residue theorem. Plotting the distribution for 1D, 2D, and 3D, we can see that the heat distribution is smoothed in the 2D and 3D case, a feature also present in the free particle.  Moreover, the asymptotic behavior is analytically derived for all dimensions and reduces to an exponential dependence in $Q$ in 1D, 2D, and 3D, a behavior predicted in \cite{rosinberg_heat_2016}. Moreover, for the 1D, this long times asymptotic behavior was recently in evidence in the literature where the exponential dependence on the modulus of $Q$ was found \cite{Fogedby_2020}.

\subsection{Linear}

\begin{figure}
    \centering
    \includegraphics[width=17cm]{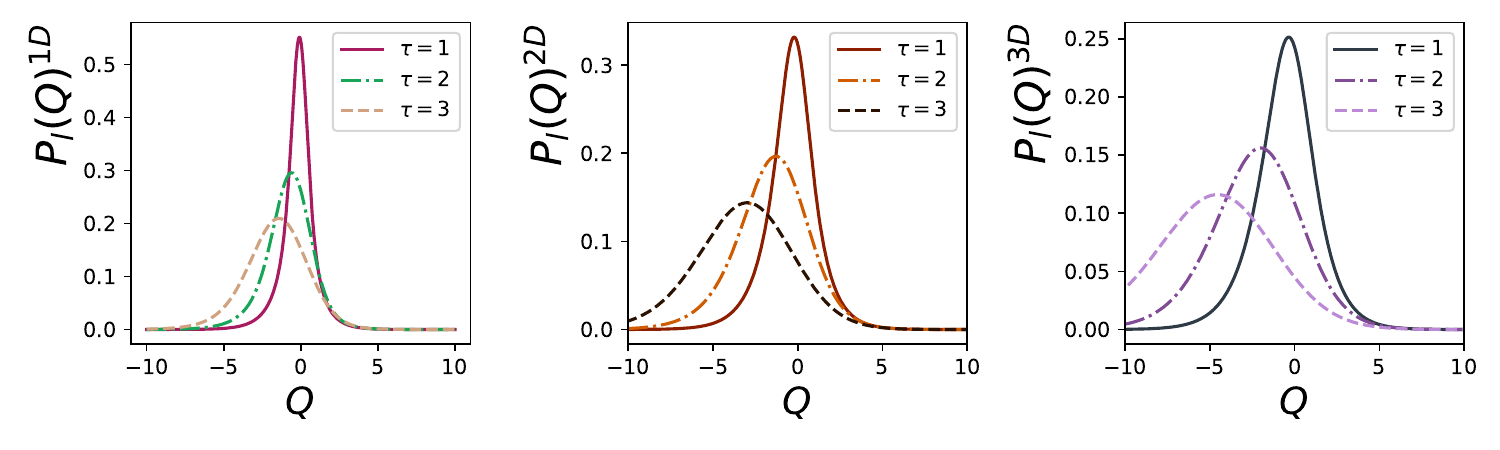}
    \caption{Heat distribution for the linear case in 1, 2, and 3 dimensions. All the remaining parameters are set to one. The behavior of the heat distribution does not change considerably as we increase the dimensions. Nevertheless, the distribution is broader in the 3D case than in the 1D and 2D cases.}
    \label{linear}
\end{figure}

The linear case, for all dimensions, needs to be solved numerically, and the result for 1D, 2D, and 3D are plotted in figure \ref{linear}. Some general considerations can be made about the heat distribution for linear potentials. With a linear potential, we break the symmetry found previously for the free particle case. Qualitatively, we can see in figure~\ref{linear} the behavior of the distribution. The distribution $P(Q)$ starts with a peak located in zero, however, as time pass the distribution starts to spread towards the negative values of $Q$. This means that it will be more probable for the particle to lose energy from the bath as time pass,   as the work done by the external force is transformed into heat that is dumped in the thermal reservoir. A reasonable interpretation of the linear potential is that the  Brownian particle has a charge and lies inside a thermal bath contained between the plates of a capacitor, and the force is the one generated by the uniform electric field.  It is straightforward to observe that the particle loses energy to the bath due to the dragging action of the electric field. In this case, we were not able to solve the integration analytically. However, we managed to get a numerically exact result through numerical integration of the characteristic function Eq.~(\ref{zlin}). Coincidentally, this case has the same structure as the overdamped ``parabola sliding system" \cite{saha_work_2014}. The derived distributions for both systems, are written in terms of integrals in $\lambda$ that are closely related. Comparing cases free and linear, with the respective similar overdamped systems, put in evidence an analogy between the underdamped and the overdamped case for these systems. The drift term in the underdamped cases fulfills the role of the harmonic force for the overdamped case, yielding mathematically similar results.

\begin{figure}
    \centering
    \includegraphics[width = 15 cm]{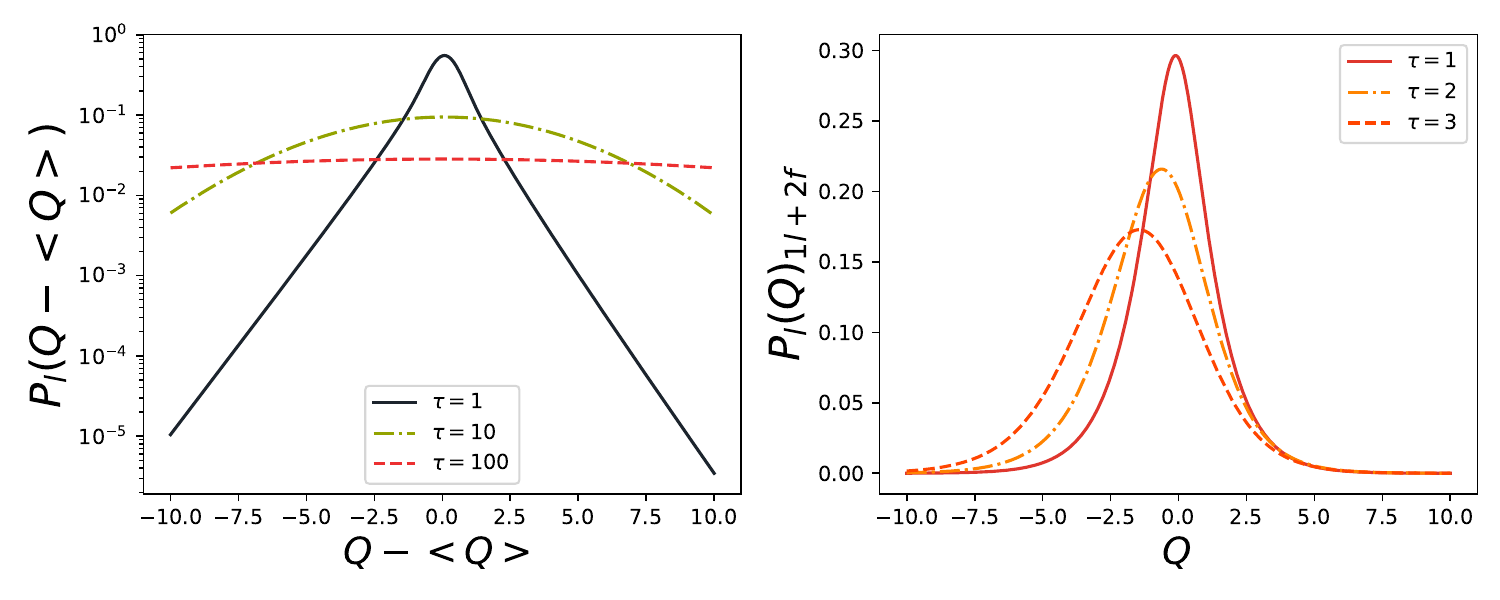}
    \caption{Left: Probability distribution of $Q-\langle Q\rangle$ over time. All remaining parameters are set to one. Right: Heat distribution for the one direction case. All remaining parameters are set to one. Note that the behavior of the distribution do not change by changing the direction of the force due to the isotropy of space.}
    \label{heatlin2}
\end{figure}

The heat distribution becomes flat for asymptotic time. A behavior already predicted by the averages. Now let us see if the probability distribution for the shifted heat $Q-\langle Q\rangle$ becomes also flat. The only change in the calculation is an extra exponential factor outside the characteristic function, that is
\begin{equation}
    P_l(Q-\langle Q\rangle) = \int \frac{d\lambda}{2\pi } e^{i\lambda (Q+\langle Q\rangle)}Z_l(\lambda),
\end{equation}
where we restrict our analysis for the case $n=1$, since the dimension does not affect the qualitative behavior in this case. The result is plotted on the left of figure~\ref{heatlin2}. One can see that the distribution is still flat in asymptotic time. However, now the distribution is symmetric around its average $\langle Q -\langle Q\rangle\rangle=0$.

Due to the isotropy of space, the result of $P_l(Q)$ can not depend on the direction in which the force is defined. This is why in the previous section, all the moments of the heat in the linear potential are symmetric in $\kappa$. To further investigate this question, we analyze the 3D case where the linear potential has only one component. This could correspond to a change of frame of reference, where the force components in the $x-$axis and $y-$axis are null. The characteristic function for this case will reduce to 
\begin{equation}
    Z_l(\lambda)_{1l+2f} = Z_l(\lambda)Z_f(\lambda)^2.
\end{equation}
Integrating numerically we find the heat distribution, $P_l(Q)_{1l+2f}$, which is plotted in the right of figure~\ref{heatlin2}. Note that the heat distribution has the same behavior as the distribution in figure~\ref{linear}. This serves as a consistency check, since, due to the isotropy of the space, the heat can not depend on the direction of the force of the linear potential.

\section{Comparison of the results}\label{sec5}
A qualitative comparison is desired since we are dealing with the fundamental models of Brownian motion. 
The three cases studied here are the most simple cases for the underdamped Langevin equation; the free particle, the linear potential, and the harmonic potential. These cases are related, it is logical to find some common features between them. Here, we considered the 1D, 2D, and 3D cases for each potential in detail, thus, we have on total nine probabilities distributions. Some of them are analytically derived, while others are derived only in the asymptotic limit. Moreover, in some cases, the distribution has a divergence. To compare the results, we list in the table \ref{tbl} the nine distributions and some qualitative features.

All the Fourier transform of (and thus the heat distributions associated) eq.~(\ref{zfree}), eq.~(\ref{zharm}), and eq.~(\ref{zlin}), can be written in terms of an oscillatory integral with oscillatory exponential dependence in $Q$ and a polynomial in the denominator of the integral inside a square root. Only for the free case, the heat distribution can be solved analytically for arbitrary dimensions, giving a Bessel function dependence.  While in the harmonic case, the distribution is only analytical for even dimensions or in the asymptotic time limit. And, for 2D, the free and harmonic case gives an explicit exponential dependence on $Q$. The linear case is only numerically accessible for all dimensions and differs significantly in the statistical behavior due to the lack of symmetry. Moreover, as a consistency check, the linear and harmonic cases are equivalent to the free case in the respective limits $\kappa=0$ and $k=0$. This is a necessary consistency test for our results. 
 
 A well-noted behavior is that in some cases the heat distribution has a non-smooth peak in $Q=0$. The free case in 1D and 2D, and the harmonic case in 1D show this behavior. While we can interpret the probability smoothing, as we increase the dimension, as the product of increasing the degrees of freedom, the non-smooth peak in low dimension harmonic and free cases can be understood in the same way that we understand the peak in the delta distribution of Dirac at the overdamped limit: it is extremely rare to have deviations from $ Q = 0 $ for the one-dimensional case. This means that the particle is on average close to equilibrium with the environment. Moreover, the results obtained for 1D, 2D, and 3D suggest that as we increase the dimension the distribution is smoothed.
 
The unknown functions that we solve numerically have very similar forms to the integral definition of the Bessel function case. Let us note, then, whether we have analytical results or not depends on the categorization of these functions. In the linear case, the difference is significant due to the dependence on $ \lambda $ on $ f_2 (\lambda,\kappa) $. However, in the harmonic case, we have a fourth-order polynomial in the denominator. We could then ask ourselves if this form of integral could not be written using generalized functions such as the Meijer-G function or others \cite{mathai1993handbook}. This type of integral is common in thermodynamic functional calculations, so it would be interesting to investigate a closed-form. Moreover, it can be seen as a generalization of the Besselian probability distribution of the free case.

\begin{table}
\caption{\label{tabone} List of results and its common qualitative features.} 
\begin{tabular}{|l|c|c|c|}
Cases: & Analytical: & Non smooth peak: & Symmetric: \\
Free & All cases  &1D and 2D & All cases \\
Linear & --- &  --- & All cases \\
Harmonic& Even dimensions/Asymptotic time& 1D & All cases
\end{tabular}\label{tbl}
\end{table}

\section{Conclusion}\label{sec6}

In conclusion, we have investigated the heat distribution in the underdamped regime for three different cases: the free particle, the constant force, and the harmonic oscillator in arbitrary dimensions. These are the most fundamental cases of Brownian motion. We review the Stochastic Thermodynamics of such systems and through the use of the path integral formalism of the stochastic processes. We were able to find analytical results for the characteristic function, its central moments, and exact results for the heat distributions for the free, harmonic, and linear cases. The harmonic and free cases are very closely related since there is no asymmetry in the potential. While the linear case can be reduced to the free case by a coordinate transformation, the heat distribution is evolving to a flat distribution. Therefore, we investigate the linear case by considering the distribution of $Q-\langle Q \rangle$, and the distribution for the 3D case with the linear potential with only one component. The work done here helps to understand the thermodynamics of fundamental systems of Brownian motion. For dimensions larger than one, the physical situations are closer to reality. Regardless of the more complicated cases studied in the literature, we thought it necessary to turn our attention to the simplest cases. Therefore, this work deals with the basis of the study of the thermodynamics of Brownian systems. In particular, the heat of the original Langevin equation has never been studied in detail, and the heat distribution of the harmonic oscillator was only studied for the equilibrium regime \cite{Fogedby_2020}, the 1D with feedback protocol \cite{rosinberg_stochastic_2017,munakata_entropy_2012}, and the overdamped regime \cite{chatterjee_exact_2010}, as far as we know. Moreover, the harmonic case is important because it appears in well-known experiments as the Carnot Brownian engine \cite{martinez2016brownian} and the Stirling Brownian engine \cite{blickle2012realization}.

New directions can come from the present work, generalizing the list of results in heat distribution of Stochastic Thermodynamics. We point out that using the same methods used herein, it is possible to solve for the heat in the case of inhomogeneous viscous force, which could be the subject of future work.

\appendix

\section{Path Integral Formalism}\label{apA}
In this paper, we use the path integral formalism. In this appendix, we show the steps of the path integral formalism of stochastic process \cite{cugliandolo2017rules,cugliandolo2019building,wio2013path,chaichian2018path}, with the adequate  simplifications, for the cases studied in this article. For simplicity, we will show in detail the 1D case, and write the generalization for higher dimensions in the end.

The transition probability between $x_0,v_0$ and $x_\tau,v_\tau$, of the Brownian motion described by eq.~(\ref{genLa}) is given by the path integral \cite{onsager1953fluctuations,chaichian2018path}
\begin{equation}
    P[v_\tau,x_\tau,\tau|v_0,x_0] = \int Dx \exp\left(-\frac{1}{4\gamma T}\mathcal{A}[v,x] \right). \label{path}
\end{equation}
The stochastic thermodynamics is formulated in the Stratonovich prescription \cite{bo2019functionals}. For most of the cases, the stochastic action will have the structure 
\begin{equation}
    \mathcal{A}[\eta]=\int_0^\tau\eta(t)^2dt,
\end{equation}
where now, the bath force $\eta$ can be seen as a ``functional of the position and velocity", since, from eq.~(\ref{genLa}),  
$\eta\rightarrow\eta[x]=m\ddot{x}+\gamma \dot{x} +\partial_x V(x,t)$. In the Stratonovich prescription of path integral \cite{bo2019functionals} there is a missing extra term in this action (see \cite{rosinberg_heat_2016}). Here we choose to omit this term since, for quadratic actions, this missing term just gives a trivial dependence on the constants of the problem. Thus, it can be absorbed into a normalization constant.

In the present paper, we only deal with quadratic stochastic actions. In this case, to solve the path integral in eq.~(\ref{path}) we can use the weak noise expansion \cite{moreno2019conditional}, which here will give the exact result. The weak noise expansion is an analog of the WKB method \cite{feynman2010quantum, bender1999advanced} used in quantum mechanics and in many other physical problems. We start by splitting  the position of the particle into a fluctuation path and a deterministic path as $x(t)=x_c(t)+y(t)$, where $x_c(t)$ have the same boundary conditions of $x(t)$, and is the solution of the extremization $\delta\mathcal{A}=0$. Making this transformation in the action, we find
\begin{equation}
    P[v_\tau,x_\tau,\tau|v_0,x_0]= \exp\left({-\frac{1}{4\gamma T}\mathcal{A}[v_c,x_c]}\right) \int Dy\; e^{ \mathcal{A}[y]}.
\end{equation}
Instead of solving the path integral in $y$, we can just use the normalization property of the transitional probability, that is
\begin{equation}
\int dv_\tau dx_\tau P[v_\tau,x_\tau,\tau|v_0,x_0,0] = 1.
\end{equation}
This gives the constraint
\begin{equation}
    \int Dy\; e^{\mathcal{A}_y} = \left(\int dv_\tau dx_\tau\exp\left({-\frac{1}{4\gamma T}\mathcal{A}[v_c,x_c]}\right)\right)^{-1}.
\end{equation}
This happens because $y(t)$ does not depend on the boundary constants $v_\tau,x_\tau,v_0,x_0$. It is important to notice that this simplification only occurs because we have a quadratic action. For more general systems this method only gives us  an approximation, and the path integral in $ y(t)$ needs to be solved \cite{moreno2019conditional}. The transitional probability will be 
\begin{equation}
    P[v_\tau,x_\tau,\tau|v_0,x_0,0] = \frac{\exp\left({-\frac{1}{4\gamma T}\mathcal{A}[v_c,x_c]}\right)}{\int dv_\tau dx_\tau\exp\left({-\frac{1}{4\gamma T}\mathcal{A}[v_c,x_c]}\right)}\label{eq:prob}
\end{equation}

For three or two dimensions, the generalization is straightforward. Since the noise is uncorrelated in its components, the joint conditional probability will be just
\begin{equation}
    P[\vec{v}_\tau,\vec{r}_\tau,\tau|\vec{v}_0,\vec{r}_0,0] = \frac{\exp\left({-\frac{1}{4\gamma T}\mathcal{A}[\vec{v}_c,\vec{r}_c]}\right)}{\int d^3\vec{v}_\tau d^3\vec{r}_\tau\exp\left({-\frac{1}{4\gamma T}\mathcal{A}[\vec{v}_c,\vec{r}_c]}\right)}\label{3dprob}
\end{equation}
we now integrate over $d^3\vec{v}_\tau d^3\vec{r}_\tau$ and the stochastic action depends on the components. Moreover, for cases where the potential can be written as a sum of the components, i.e. is separable by addition, joint conditional probability will be the product of the individual conditional probabilities for the $x,y,z$ components. That is, for 3D we have
\begin{equation}
    P[\vec{v}_\tau,\vec{r}_\tau,\tau|\vec{v}_0,\vec{r}_0]= P[v_{x,\tau},x_\tau,\tau|v_{x,0},x_0]P[v_{y,\tau},y_\tau,\tau|v_{y,0},y_0]P[v_{z,\tau},z_\tau,\tau|v_{z,0},z_0].
\end{equation}
In this case, the contribution from each component is the same. Therefore, we only need to calculate one of the contributions and use it to find the others.

\section{Path Integral and characteristic function harmonic case}\label{apB}

The stochastic action of the harmonic potential case will be
\begin{equation}
    \mathcal{A}[x]= - \frac{1}{4\gamma T}\int_0^\tau \left(m\ddot x + \gamma \dot{x} +  k x\right)^2 dt.
\end{equation}
We need to extremize the stochastic action, obtaining, 
\begin{eqnarray}
     m^2 x^{(4)}(t) +(2km- \gamma ^2) \ddot{x}(t)+ k^2 x(t)=0,
\end{eqnarray}
where this equation has four boundary conditions $v_0, x_0, v_\tau, x_\tau$. The extreme solution is,
\begin{equation}
     x_c(t)= d_1\, {\rm exp}\left({\frac{-\Psi^+ t}{2 m}} \right)+ d_2 \, {\rm exp}\left(\frac{-\Psi^- t}{2 m}\right) + d_3 \, {\rm exp}\left(\frac{\Psi^- t}{2 m}\right)+ d_4 \, {\rm exp}\left(\frac{\Psi^+ t}{2 m}\right),
\end{equation}
with $\Psi^+ = \gamma + \sqrt{\gamma^2 - 4 k m}$ and $\Psi^- = \gamma - \sqrt{\gamma^2 -4 k m}$. Now we will use the boundary conditions $x_c(0)=x_0$, $x_c(\tau)=x_\tau$,  $\dot{x}_c(0)=v_0$ and $\dot{x}_c(\tau)=v_\tau$ to compute the constants $d$'s. The extreme action reads,
\begin{eqnarray}
     \mathcal{A}[x_c] = \frac{1}{4 m T}\Bigg[ \gamma \, d_3^2 \, \Psi^- \left( e^{\frac{
    \Psi^- \tau}{m}} -1\right) + 8 \, d_3\,d_4 \, k\, &m& \left(e^{\frac{\gamma  \tau }{m}}-1\right)  +\gamma \, d_4^2 \,\Psi^+ \left(e^{\frac{\Psi^+ \tau}{m}}-1\right)\Bigg]
\end{eqnarray}
with
\begin{eqnarray}
     d_3 = \frac{e^{-\frac{\, \Psi^+\tau}{2 m}} }{4\Delta}\Bigg(\gamma  \, e^{\frac{\left(3\Psi^+ -2\gamma \right)\tau}{2 m}} \Big(\, x_0 \Psi^+  + 2 m v_0 \Big) -2\, m\, e^{\frac{\, \Psi^+\tau}{2 m}} \Big(v_0 \, \Psi^+ +2 k x_0\Big)\nonumber \\
    +e^{-\frac{\Psi^- \tau}{2 m}} \Big(2 m v_0 \sqrt{\gamma^2 - 4 k m} -\gamma  x_0 \, \Psi^+ +4 k m x_0\Big)-2 m \,e^{\frac{\sqrt{\gamma^2 - 4 k m}\tau}{m}} \Big(v_\tau \, \Psi^+ +2 k x_\tau \Big)  +\nonumber\\
     +e^{\frac{\, \Psi^+\tau}{m}} \Big(2 m v_\tau \sqrt{\gamma^2 - 4 k m} - \gamma  x_\tau \, \Psi^+ +4 k m x_\tau\Big)+\gamma  \Big(x_\tau \, \Psi^+ +2 m v_\tau\Big)
    \Bigg)
\end{eqnarray}

\begin{eqnarray}
 d_4 = \frac{e^{-\frac{\tau\sqrt{\gamma^2 - 4 k m}}{m}}}{4\Delta}\Bigg(  \gamma e^{-\frac{\left(3\Psi^- -2\gamma\right)\tau}{2m}}  (x_\tau \Psi^- + 2 m v_\tau) + \gamma(x_0 \Psi^- + 2 m v_0)+  \nonumber\\
-2\, m\, e^{\frac{-\Psi^- \tau}{2m}}(v_\tau \Psi^- +2 k x_\tau)  + e^{\frac{-\Psi^- \tau}{m}} (4 k m x_0 -2 m v_0\sqrt{\gamma^2 - 4 k m}-\gamma x_0 \Psi_-) + \nonumber \\
 e^{\frac{\Psi^+ \tau}{m}}(4 k m x_\tau - 2 m v_\tau\sqrt{\gamma^2 - 4 k m } - \gamma x_\tau \Psi^-)-2\, m\, e^{\frac{\tau\sqrt{\gamma^2 - 4 k m}}{m}}(v_0 \Psi^- +2 k x_0)\Bigg)
\end{eqnarray}
And, in both constants, the factor $\Delta$ is,
\begin{equation}
    \Delta = -4 k m + (4 k m -\gamma^2)\cosh\left(\frac{\gamma \tau}{m}\right)+\gamma^2 \cosh\left( \frac{\tau \sqrt{\gamma^2 - 4 k m}}{m} \right).
\end{equation}
Although in the extreme action we have long constants terms, due to the large set of parameters involved in our specific model, we can rewrite $\mathcal{A}[x_c]$ in a bilinear form in terms of vector $\vec{b} = (v_\tau,v_0,x_\tau,x_0)$. So, accordingly to eq.~(\ref{eq:prob}), the normalized path integral is,
\begin{equation}
    \frac{\exp\left({-\frac{1}{4\gamma T}\mathcal{A}[v_c,x_c]}\right)}{\int dv_\tau dx_\tau\exp\left({-\frac{1}{4\gamma T}\mathcal{A}[v_c,x_c]}\right)} =  N' \exp\left(-\frac{1}{D}\vec{b}\;\mathcal{M}\;\vec{b}^T\right)\label{pathharmo}
\end{equation}
where
\begin{equation}
    N'=\frac{ \sqrt{k m \left(4 k m-\gamma ^2\right)} e^{\frac{\gamma  \tau }{2 m}}}{2 \sqrt{2} \pi  T \sqrt{\gamma ^2 \cosh \left(\frac{\tau  \sqrt{\gamma ^2-4 k m}}{m}\right)+\left(4 k m-\gamma ^2\right) \cosh \left(\frac{\gamma  \tau }{m}\right)-4 k m}}
\end{equation}
and
\begin{equation}
 \mathcal{M} = 
\left(
\begin{array}{cccc} 
    A_1 & A_3 & A_4 & A_5 \\ 
    0 & A_1+\frac{m}{2T} & -A_5 & -A_4 \\
    0 & 0 & A_2 & A_6 \\
    0 & 0 & 0 & A_2+\frac{k}{2T}
\end{array}
\right)\label{matr}
\end{equation}
The terms $A_i$'s  in the matrix $\mathcal{M}$ contains the dependence on the constants of the problem. These terms are
\begin{eqnarray}
 \nonumber A_1=\frac{m}{4T\Delta} \left(-\gamma  \sqrt{\gamma ^2-4 k m} \sinh \left(\frac{\tau  \sqrt{\gamma ^2-4 k m}}{m}\right) +\gamma ^2 \left(-\cosh \left(\frac{\tau  \sqrt{\gamma ^2-4 k m}}{m}\right)\right)+\right.\\\left.+\left(\gamma ^2-4 k m\right)\left( \cosh \left(\frac{\gamma  \tau }{m}\right)+\sinh \left(\frac{\gamma  \tau }{m}\right)\right)+4 k m\right),
\end{eqnarray}
\begin{equation}
 A_2 = \frac{k}{m}A_1 +\frac{k\gamma  \sqrt{\gamma ^2-4 k m} \sinh \left(\frac{\tau  \sqrt{\gamma ^2-4 k m}}{m}\right)}{2 T \Delta},
\end{equation}
\begin{eqnarray}
     \nonumber A_3 =\frac{m}{T\Delta}  \left(\left(4 k m-\gamma ^2\right) \sinh \left(\frac{\gamma  \tau }{2 m}\right) \cosh \left(\frac{\tau  \sqrt{\gamma ^2-4 k m}}{2 m}\right)+\right.\\\left.+\gamma  \sqrt{\gamma ^2-4 k m} \cosh \left(\frac{\gamma  \tau }{2 m}\right) \sinh \left(\frac{\tau  \sqrt{\gamma ^2-4 k m}}{2 m}\right)\right),
\end{eqnarray}
\begin{eqnarray}
 A_4 = -\frac{2\gamma  k m \sinh ^2\left(\frac{\tau  \sqrt{\gamma ^2-4 k m}}{2 m}\right)}{T \Delta}, \\
 A_5  = \frac{2k m \sqrt{\gamma ^2-4 k m} \sinh \left(\frac{\gamma  \tau }{2 m}\right) \sinh \left(\frac{\tau  \sqrt{\gamma ^2-4 k m}}{2 m}\right)}{T \Delta}, \\
 A_6 = \frac{k}{m} A_3 - \frac{2k\gamma  \sqrt{\gamma ^2-4 k m} \cosh \left(\frac{\gamma  \tau }{2 m}\right) \sinh \left(\frac{\tau  \sqrt{\gamma ^2-4 k m}}{2 m}\right)}{T \Delta}\;.
\end{eqnarray}

Thus, equation~\ref{pathharmo} is the transitional probability for the harmonic case. Having the transitional probability we can calculate the characteristic function and consequently the heat distribution. Despite its complicated formula, the transitional probability is a Gaussian probability, and using an equilibrium initial distribution, the characteristic function will be given by the Gaussian integral
\begin{equation}
    Z_h(\lambda)=\mathcal{N}'  \int d^4\vec{b}\, P_0(v_0,x_0)\, \exp\left(-\vec{b}\;\mathcal{D}(\lambda)\;\vec{b}^T\right)
\end{equation}
where $\mathcal{D}(\lambda)$ is
\begin{equation}
    \mathcal{D}(\lambda)= \mathcal{M} + \frac{i\lambda}{2}\textbf{diag}\left(-m,m,-k,k\right),
\end{equation}
where $\textbf{diag}$ means that we have a 4 dimensional diagonal matrix, with elements given in the argument. The above integral is just a Gaussian integral, and the result is given in equation~\ref{zharm}. Note that, here we only derive the 1D result, because as mentioned early, to higher dimensions the generalization is trivial.

\section{Path Integral and characteristic function linear case}\label{apC}

The stochastic action for the linear case is
\begin{equation}
    \mathcal{A}[x(t)]= \int_0^\tau \left(m\ddot{x}+\gamma\dot{x}-\kappa\right)^2 dt
\end{equation}
The extremization of this action will give the solution $x_c(t)$,
\begin{equation}
    \delta \mathcal{A} = 0 \rightarrow  m^2 x_c^{(4)}(t)= \gamma ^2 \ddot{x}_c(t), \label{c2}
\end{equation}
where the superscript $(4)$ denote 4-th order time derivatives. Solving the above differential equation we find
\begin{equation}
    x_c(t)= \frac{c_1 m^2 e^{\frac{\gamma  t}{m}}}{\gamma ^2}+\frac{c_2 m^2 e^{-\frac{\gamma  t}{m}}}{\gamma ^2}+c_4 t+c_3
\end{equation}
where the $c's$ are arbitrary constant, which needs to be fixed by the equations: $x_c(0)=x_0,x_c(\tau)=x_\tau,\dot{x}_c(0)=v_0,\dot{x}_c(\tau)=v_\tau$. Note that, since we have a fourth derivative in equation~\ref{c2} we these four boundary conditions. The action evaluated in $x_c$ will be
\begin{eqnarray}
    \mathcal{A}[x_c]= \gamma ^2 c_4{}^2 \tau -2 \gamma  c_4 \kappa  \tau +\kappa ^2 \tau +4 c_1 c_4 m^2 \left(e^{\frac{\gamma  \tau }{m}}-1\right)+\\+\frac{2 c_1 m^2 }{\gamma }\left(e^{\frac{\gamma  \tau }{m}}-1\right) \left(-2 \kappa +c_1 m \left(e^{\frac{\gamma  \tau }{m}}+1\right)\right) \label{actioncase2}
\end{eqnarray}
where 
\begin{eqnarray}
  c_1= \frac{\gamma}{2m}\left(\frac{v_\tau-v_0}{e^{\frac{\gamma  \tau }{m}}-1}+\frac{\gamma  (\tau  (v_0+v_\tau)+2 x_0-2 x_\tau)}{\gamma  \tau +e^{\frac{\gamma  \tau }{m}} (\gamma  \tau -2 m)+2 m}\right), \\  c_4 = \frac{m (v_0+v_\tau) \left(e^{\frac{\gamma  \tau }{m}}-1\right)+\gamma  (x_0-x_\tau) \left(e^{\frac{\gamma  \tau }{m}}+1\right)}{-\gamma  \tau +e^{\frac{\gamma  \tau }{m}} (2 m-\gamma  \tau )-2 m}.
\end{eqnarray}
Substituting these constants, we have the path integral for this action giving a quadratic dependence in the boundary terms, that is
\begin{eqnarray}
  \int Dx \exp\left(-\frac{1}{4\gamma T}\int_0^\tau \left(m\ddot{x}+\gamma \dot{x} -\kappa\right)^2dt\right) =\mathcal{N}\exp{\left(f_1(v_\tau,v_0,x_\tau,x_0)\right)}\label{pi2}
\end{eqnarray}
where $f_1$ and $\mathcal{N}$ are
\begin{eqnarray}
  \nonumber f_1(v_\tau,v_0,x_\tau,x_0)=-\frac{1}{8 \gamma  T \left(\gamma  \tau  \coth \left(\frac{\gamma  \tau }{2 m}\right)-2 m\right)}\times \\\nonumber \Bigg[\left.m (\gamma  (v_0+v_\tau)-2 \kappa ) (2 \kappa  \tau +4 m (v_0-v_\tau)+\gamma  \tau  (v_0+v_\tau)+4 \gamma  (x_0-x_\tau))\right. + \\\nonumber +\gamma  \coth \left(\frac{\gamma  \tau }{2 m}\right) \left(2 m^2 (v_0-v_\tau)^2+2 m \tau  (v_0-v_\tau) (\gamma  (v_0+v_\tau)-2 \kappa )+\right.\\\left.\left.-\gamma  m \tau  (v_0-v_\tau)^2 \coth \left(\frac{\gamma  \tau }{2 m}\right)-2 (\kappa  \tau +\gamma  x_0-\gamma  x_\tau)^2\right) \right],  \label{f1}
\end{eqnarray}
\begin{equation}
    \mathcal{N}=\frac{\gamma }{4 \pi  T}\frac{\sqrt{m} e^{\frac{\gamma  \tau }{2 m}}}{\sqrt{\gamma  \tau  \sinh \left(\frac{\gamma  \tau }{m}\right)-2 m \cosh \left(\frac{\gamma  \tau }{m}\right)+2 m}}.
\end{equation}
The dependence of the boundary conditions is written in the term $f_1$, and is a quadratic polynomial in the boundary terms. 

The characteristic function (for 1D) case yields a Gaussian integral over $x_0,x_\tau,v_0,v_\tau$. Defining $\Gamma_{(0,\tau)}=\{x_{(0,\tau)},v_{(0,\tau)}\}$ as the boundary terms. First, the integral in $x_\tau$ in the characteristic function gives (ignoring the multiplicative terms)
\begin{eqnarray}
   Z_l(\lambda)= \int d\Gamma_0 dv_\tau \int dx_\tau  \exp\left[f_1\left(\Gamma_t,\Gamma_0\right) + i\lambda \left(\kappa(x_\tau-x_0)-\frac{m}{2}(v_\tau^2-v_0^2)\right)\right] \sim \\\nonumber \sim  \int d\Gamma_0 dv_\tau\exp{\left(\frac{\tanh \left(\frac{\gamma  \tau }{2 m}\right) \left(\gamma  \kappa  \tau  (1+2 i \lambda  T) \coth \left(\frac{\gamma  \tau }{2 m}\right)+\kappa  m (-2-4 i \lambda  T)+\gamma  m (v_0+v_\tau)\right)^2}{4 \gamma ^2 T \left(\gamma  \tau  \coth \left(\frac{\gamma  \tau }{2 m}\right)-2 m\right)}\right)}
\end{eqnarray}
At this point, after integrating in $x_\tau$, we notice that there is no 
dependence in $x_0$ due to translation symmetry. This means that we can neglect the integration in 
$x_0$, just using  that the probability  $P(x_0,v_0)$ satisfies $\int P(x_0,v_0)\, dx_0 = P(v_0)$. Assuming an initially thermalized  distribution for the velocity, we have
\begin{equation}
    P(v_0)  = \sqrt{\frac{m}{2\pi T}}\exp\left(-\frac{1}{2T}mv_0^2\right).
\end{equation}
The subsequent integration in $v_\tau,v_0$ will be straightforward, since it is just a Gaussian integral. Integrating in $v_\tau$ and $v_0$, and collecting the multiplicative terms, gives the correct result for the characteristic function in equation~\ref{zlin}. Where the function $f_2(\lambda,\kappa)$ is given by
\begin{eqnarray}
   f_2(\lambda,\kappa) = \frac{1}{\gamma ^2 \left(\cosh \left(\frac{\gamma  \tau }{m}\right)+\left(2 \lambda ^2 T^2+1\right) \sinh \left(\frac{\gamma  \tau }{m}\right)\right)} \times \\ \nonumber \left[ 2 \kappa ^2 \lambda ^4 T^3 \left(2 m \left(\cosh \left(\frac{\gamma  \tau }{m}\right)-1\right)-\gamma  \tau  \sinh \left(\frac{\gamma  \tau }{m}\right)\right) +\right. 2 i \kappa ^2 \lambda ^3 T^2 \left(\gamma  \tau  \sinh \left(\frac{\gamma  \tau }{m}\right)-2 m \cosh \left(\frac{\gamma  \tau }{m}\right)+2 m\right)+ \\ \nonumber -i \kappa ^2 \lambda  \left((m-\gamma  \tau ) \sinh \left(\frac{\gamma  \tau }{m}\right)+(2 m-\gamma  \tau ) \cosh \left(\frac{\gamma  \tau }{m}\right)-2 m\right)+ \\\left. +\nonumber \kappa ^2 \lambda ^2 T \left((m-\gamma  \tau ) \sinh \left(\frac{\gamma  \tau }{m}\right)+(2 m-\gamma  \tau ) \cosh \left(\frac{\gamma  \tau }{m}\right)-2 m\right)\right].
\end{eqnarray}

\bibliography{name}

\begin{thebibliography}{65}
\expandafter\ifx\csname natexlab\endcsname\relax\def\natexlab#1{#1}\fi
\expandafter\ifx\csname bibnamefont\endcsname\relax
  \def\bibnamefont#1{#1}\fi
\expandafter\ifx\csname bibfnamefont\endcsname\relax
  \def\bibfnamefont#1{#1}\fi
\expandafter\ifx\csname citenamefont\endcsname\relax
  \def\citenamefont#1{#1}\fi
\expandafter\ifx\csname url\endcsname\relax
  \def\url#1{\texttt{#1}}\fi
\expandafter\ifx\csname urlprefix\endcsname\relax\def\urlprefix{URL }\fi
\providecommand{\bibinfo}[2]{#2}
\providecommand{\eprint}[2][]{\url{#2}}

\bibitem[{\citenamefont{Oliveira}(2020)}]{oliveira2020classical}
\bibinfo{author}{\bibfnamefont{M.~J.~d.} \bibnamefont{Oliveira}},
  \bibinfo{journal}{Revista Brasileira de Ensino de F{\'\i}sica}
  \textbf{\bibinfo{volume}{42}} (\bibinfo{year}{2020}).

\bibitem[{\citenamefont{Ciliberto}(2017)}]{ciliberto_experiments_2017}
\bibinfo{author}{\bibfnamefont{S.}~\bibnamefont{Ciliberto}},
  \bibinfo{journal}{Phys. Rev. X} \textbf{\bibinfo{volume}{7}},
  \bibinfo{pages}{021051} (\bibinfo{year}{2017}), \bibinfo{note}{publisher:
  American Physical Society},
  \urlprefix\url{https://link.aps.org/doi/10.1103/PhysRevX.7.021051}.

\bibitem[{\citenamefont{Sekimoto}(2010)}]{sekimoto2010stochastic}
\bibinfo{author}{\bibfnamefont{K.}~\bibnamefont{Sekimoto}},
  \emph{\bibinfo{title}{Stochastic energetics}}, vol. \bibinfo{volume}{799}
  (\bibinfo{publisher}{Springer}, \bibinfo{year}{2010}).

\bibitem[{\citenamefont{Ryabov}(2015)}]{ryabov2015stochastic}
\bibinfo{author}{\bibfnamefont{A.}~\bibnamefont{Ryabov}},
  \emph{\bibinfo{title}{Stochastic dynamics and energetics of biomolecular
  systems}} (\bibinfo{publisher}{Springer}, \bibinfo{year}{2015}).

\bibitem[{\citenamefont{Seifert}(2019)}]{seifert_stochastic_2019}
\bibinfo{author}{\bibfnamefont{U.}~\bibnamefont{Seifert}},
  \bibinfo{journal}{Annual Review of Condensed Matter Physics}
  \textbf{\bibinfo{volume}{10}}, \bibinfo{pages}{171} (\bibinfo{year}{2019}),
  \bibinfo{note}{\_eprint:
  https://doi.org/10.1146/annurev-conmatphys-031218-013554},
  \urlprefix\url{https://doi.org/10.1146/annurev-conmatphys-031218-013554}.

\bibitem[{\citenamefont{Peliti and Pigolotti}(2021)}]{peliti2021stochastic}
\bibinfo{author}{\bibfnamefont{L.}~\bibnamefont{Peliti}} \bibnamefont{and}
  \bibinfo{author}{\bibfnamefont{S.}~\bibnamefont{Pigolotti}},
  \emph{\bibinfo{title}{Stochastic Thermodynamics: An Introduction}}
  (\bibinfo{publisher}{Princeton University Press}, \bibinfo{year}{2021}).

\bibitem[{\citenamefont{Seifert}(2005)}]{seifert2005entropy}
\bibinfo{author}{\bibfnamefont{U.}~\bibnamefont{Seifert}},
  \bibinfo{journal}{Physical review letters} \textbf{\bibinfo{volume}{95}},
  \bibinfo{pages}{040602} (\bibinfo{year}{2005}).

\bibitem[{\citenamefont{Chernyak et~al.}(2006)\citenamefont{Chernyak, Chertkov,
  and Jarzynski}}]{chernyak2006path}
\bibinfo{author}{\bibfnamefont{V.~Y.} \bibnamefont{Chernyak}},
  \bibinfo{author}{\bibfnamefont{M.}~\bibnamefont{Chertkov}}, \bibnamefont{and}
  \bibinfo{author}{\bibfnamefont{C.}~\bibnamefont{Jarzynski}},
  \bibinfo{journal}{Journal of Statistical Mechanics: Theory and Experiment}
  \textbf{\bibinfo{volume}{2006}}, \bibinfo{pages}{P08001}
  (\bibinfo{year}{2006}).

\bibitem[{\citenamefont{Jarzynski}(2011)}]{jarzynski2011equalities}
\bibinfo{author}{\bibfnamefont{C.}~\bibnamefont{Jarzynski}},
  \bibinfo{journal}{Annu. Rev. Condens. Matter Phys.}
  \textbf{\bibinfo{volume}{2}}, \bibinfo{pages}{329} (\bibinfo{year}{2011}).

\bibitem[{\citenamefont{Gong and Quan}(2015)}]{gong2015jarzynski}
\bibinfo{author}{\bibfnamefont{Z.}~\bibnamefont{Gong}} \bibnamefont{and}
  \bibinfo{author}{\bibfnamefont{H.}~\bibnamefont{Quan}},
  \bibinfo{journal}{Physical Review E} \textbf{\bibinfo{volume}{92}},
  \bibinfo{pages}{012131} (\bibinfo{year}{2015}).

\bibitem[{\citenamefont{Seifert}(2012)}]{seifert2012stochastic}
\bibinfo{author}{\bibfnamefont{U.}~\bibnamefont{Seifert}},
  \bibinfo{journal}{Reports on progress in physics}
  \textbf{\bibinfo{volume}{75}}, \bibinfo{pages}{126001}
  (\bibinfo{year}{2012}).

\bibitem[{\citenamefont{Paraguass{\'u} and
  Morgado}(2021)}]{paraguassu_heat_2021}
\bibinfo{author}{\bibfnamefont{P.~V.} \bibnamefont{Paraguass{\'u}}}
  \bibnamefont{and} \bibinfo{author}{\bibfnamefont{W.~A.~M.}
  \bibnamefont{Morgado}}, \bibinfo{journal}{J. Stat. Mech.}
  \textbf{\bibinfo{volume}{2021}}, \bibinfo{pages}{023205}
  (\bibinfo{year}{2021}), ISSN \bibinfo{issn}{1742-5468},
  \bibinfo{note}{publisher: IOP Publishing},
  \urlprefix\url{https://doi.org/10.1088/1742-5468/abda25}.

\bibitem[{\citenamefont{Gupta and Sivak}(2021)}]{gupta_heat_2021}
\bibinfo{author}{\bibfnamefont{D.}~\bibnamefont{Gupta}} \bibnamefont{and}
  \bibinfo{author}{\bibfnamefont{D.~A.} \bibnamefont{Sivak}},
  \bibinfo{journal}{arXiv:2103.09358 [cond-mat]}  (\bibinfo{year}{2021}),
  \bibinfo{note}{arXiv: 2103.09358},
  \urlprefix\url{http://arxiv.org/abs/2103.09358}.

\bibitem[{\citenamefont{Fogedby}(2020{\natexlab{a}})}]{fogedby_heat_2020}
\bibinfo{author}{\bibfnamefont{H.~C.} \bibnamefont{Fogedby}},
  \bibinfo{journal}{J. Stat. Mech.} \textbf{\bibinfo{volume}{2020}},
  \bibinfo{pages}{083208} (\bibinfo{year}{2020}{\natexlab{a}}), ISSN
  \bibinfo{issn}{1742-5468}, \bibinfo{note}{publisher: IOP Publishing},
  \urlprefix\url{https://doi.org/10.1088/1742-5468/aba7b2}.

\bibitem[{\citenamefont{Goswami}(2019)}]{goswami_heat_2019}
\bibinfo{author}{\bibfnamefont{K.}~\bibnamefont{Goswami}},
  \bibinfo{journal}{Phys. Rev. E} \textbf{\bibinfo{volume}{99}},
  \bibinfo{pages}{012112} (\bibinfo{year}{2019}), \bibinfo{note}{publisher:
  American Physical Society},
  \urlprefix\url{https://link.aps.org/doi/10.1103/PhysRevE.99.012112}.

\bibitem[{\citenamefont{Crisanti et~al.}(2017)\citenamefont{Crisanti,
  Sarracino, and Zannetti}}]{crisanti_heat_2017}
\bibinfo{author}{\bibfnamefont{A.}~\bibnamefont{Crisanti}},
  \bibinfo{author}{\bibfnamefont{A.}~\bibnamefont{Sarracino}},
  \bibnamefont{and} \bibinfo{author}{\bibfnamefont{M.}~\bibnamefont{Zannetti}},
  \bibinfo{journal}{Phys. Rev. E} \textbf{\bibinfo{volume}{95}},
  \bibinfo{pages}{052138} (\bibinfo{year}{2017}), \bibinfo{note}{publisher:
  American Physical Society},
  \urlprefix\url{https://link.aps.org/doi/10.1103/PhysRevE.95.052138}.

\bibitem[{\citenamefont{Ghosal and Cherayil}(2016)}]{ghosal_distribution_2016}
\bibinfo{author}{\bibfnamefont{A.}~\bibnamefont{Ghosal}} \bibnamefont{and}
  \bibinfo{author}{\bibfnamefont{B.~J.} \bibnamefont{Cherayil}},
  \bibinfo{journal}{J. Stat. Mech.} \textbf{\bibinfo{volume}{2016}},
  \bibinfo{pages}{043201} (\bibinfo{year}{2016}), ISSN
  \bibinfo{issn}{1742-5468}, \bibinfo{note}{publisher: IOP Publishing},
  \urlprefix\url{https://doi.org/10.1088/1742-5468/2016/04/043201}.

\bibitem[{\citenamefont{Rosinberg et~al.}(2016)\citenamefont{Rosinberg, Tarjus,
  and Munakata}}]{rosinberg_heat_2016}
\bibinfo{author}{\bibfnamefont{M.~L.} \bibnamefont{Rosinberg}},
  \bibinfo{author}{\bibfnamefont{G.}~\bibnamefont{Tarjus}}, \bibnamefont{and}
  \bibinfo{author}{\bibfnamefont{T.}~\bibnamefont{Munakata}},
  \bibinfo{journal}{EPL} \textbf{\bibinfo{volume}{113}}, \bibinfo{pages}{10007}
  (\bibinfo{year}{2016}), ISSN \bibinfo{issn}{0295-5075},
  \bibinfo{note}{publisher: IOP Publishing},
  \urlprefix\url{https://doi.org/10.1209/0295-5075/113/10007}.

\bibitem[{\citenamefont{Kim et~al.}(2014)\citenamefont{Kim, Kwon, and
  Park}}]{kim_heat_2014}
\bibinfo{author}{\bibfnamefont{K.}~\bibnamefont{Kim}},
  \bibinfo{author}{\bibfnamefont{C.}~\bibnamefont{Kwon}}, \bibnamefont{and}
  \bibinfo{author}{\bibfnamefont{H.}~\bibnamefont{Park}},
  \bibinfo{journal}{Phys. Rev. E} \textbf{\bibinfo{volume}{90}},
  \bibinfo{pages}{032117} (\bibinfo{year}{2014}), ISSN
  \bibinfo{issn}{1539-3755, 1550-2376},
  \urlprefix\url{https://link.aps.org/doi/10.1103/PhysRevE.90.032117}.

\bibitem[{\citenamefont{Ku{\'s}mierz et~al.}(2014)\citenamefont{Ku{\'s}mierz,
  Rubi, and Gudowska-Nowak}}]{kusmierz_heat_2014}
\bibinfo{author}{\bibfnamefont{{\textbackslash}.}~\bibnamefont{Ku{\'s}mierz}},
  \bibinfo{author}{\bibfnamefont{J.~M.} \bibnamefont{Rubi}}, \bibnamefont{and}
  \bibinfo{author}{\bibfnamefont{E.}~\bibnamefont{Gudowska-Nowak}},
  \bibinfo{journal}{J. Stat. Mech.} \textbf{\bibinfo{volume}{2014}},
  \bibinfo{pages}{P09002} (\bibinfo{year}{2014}), ISSN
  \bibinfo{issn}{1742-5468}, \bibinfo{note}{publisher: IOP Publishing},
  \urlprefix\url{https://doi.org/10.1088/1742-5468/2014/09/p09002}.

\bibitem[{\citenamefont{Saha and Mukherji}(2014)}]{saha_work_2014}
\bibinfo{author}{\bibfnamefont{B.}~\bibnamefont{Saha}} \bibnamefont{and}
  \bibinfo{author}{\bibfnamefont{S.}~\bibnamefont{Mukherji}},
  \bibinfo{journal}{J. Stat. Mech.} \textbf{\bibinfo{volume}{2014}},
  \bibinfo{pages}{P08014} (\bibinfo{year}{2014}), ISSN
  \bibinfo{issn}{1742-5468},
  \urlprefix\url{https://iopscience.iop.org/article/10.1088/1742-5468/2014/08/P08014}.

\bibitem[{\citenamefont{Chatterjee and
  Cherayil}(2011)}]{chatterjee_single-molecule_2011}
\bibinfo{author}{\bibfnamefont{D.}~\bibnamefont{Chatterjee}} \bibnamefont{and}
  \bibinfo{author}{\bibfnamefont{B.~J.} \bibnamefont{Cherayil}},
  \bibinfo{journal}{J. Stat. Mech.} \textbf{\bibinfo{volume}{2011}},
  \bibinfo{pages}{P03010} (\bibinfo{year}{2011}), ISSN
  \bibinfo{issn}{1742-5468},
  \urlprefix\url{https://iopscience.iop.org/article/10.1088/1742-5468/2011/03/P03010}.

\bibitem[{\citenamefont{Chatterjee and Cherayil}(2010)}]{chatterjee_exact_2010}
\bibinfo{author}{\bibfnamefont{D.}~\bibnamefont{Chatterjee}} \bibnamefont{and}
  \bibinfo{author}{\bibfnamefont{B.~J.} \bibnamefont{Cherayil}},
  \bibinfo{journal}{Phys. Rev. E} \textbf{\bibinfo{volume}{82}},
  \bibinfo{pages}{051104} (\bibinfo{year}{2010}), ISSN
  \bibinfo{issn}{1539-3755, 1550-2376},
  \urlprefix\url{https://link.aps.org/doi/10.1103/PhysRevE.82.051104}.

\bibitem[{\citenamefont{Fogedby and Imparato}(2009)}]{fogedby_heat_2009}
\bibinfo{author}{\bibfnamefont{H.~C.} \bibnamefont{Fogedby}} \bibnamefont{and}
  \bibinfo{author}{\bibfnamefont{A.}~\bibnamefont{Imparato}},
  \bibinfo{journal}{J. Phys. A: Math. Theor.} \textbf{\bibinfo{volume}{42}},
  \bibinfo{pages}{475004} (\bibinfo{year}{2009}), ISSN
  \bibinfo{issn}{1751-8121}, \bibinfo{note}{publisher: IOP Publishing},
  \urlprefix\url{https://doi.org/10.1088/1751-8113/42/47/475004}.

\bibitem[{\citenamefont{Imparato et~al.}(2008)\citenamefont{Imparato, Jop,
  Petrosyan, and Ciliberto}}]{imparato_probability_2008}
\bibinfo{author}{\bibfnamefont{A.}~\bibnamefont{Imparato}},
  \bibinfo{author}{\bibfnamefont{P.}~\bibnamefont{Jop}},
  \bibinfo{author}{\bibfnamefont{A.}~\bibnamefont{Petrosyan}},
  \bibnamefont{and}
  \bibinfo{author}{\bibfnamefont{S.}~\bibnamefont{Ciliberto}},
  \bibinfo{journal}{J. Stat. Mech.} \textbf{\bibinfo{volume}{2008}},
  \bibinfo{pages}{P10017} (\bibinfo{year}{2008}), ISSN
  \bibinfo{issn}{1742-5468},
  \urlprefix\url{https://iopscience.iop.org/article/10.1088/1742-5468/2008/10/P10017}.

\bibitem[{\citenamefont{Imparato et~al.}(2007)\citenamefont{Imparato, Peliti,
  Pesce, Rusciano, and Sasso}}]{imparato_work_2007}
\bibinfo{author}{\bibfnamefont{A.}~\bibnamefont{Imparato}},
  \bibinfo{author}{\bibfnamefont{L.}~\bibnamefont{Peliti}},
  \bibinfo{author}{\bibfnamefont{G.}~\bibnamefont{Pesce}},
  \bibinfo{author}{\bibfnamefont{G.}~\bibnamefont{Rusciano}}, \bibnamefont{and}
  \bibinfo{author}{\bibfnamefont{A.}~\bibnamefont{Sasso}},
  \bibinfo{journal}{Phys. Rev. E} \textbf{\bibinfo{volume}{76}},
  \bibinfo{pages}{050101} (\bibinfo{year}{2007}), \bibinfo{note}{publisher:
  American Physical Society},
  \urlprefix\url{https://link.aps.org/doi/10.1103/PhysRevE.76.050101}.

\bibitem[{\citenamefont{Joubaud et~al.}(2007)\citenamefont{Joubaud, Garnier,
  and Ciliberto}}]{joubaud_fluctuation_2007}
\bibinfo{author}{\bibfnamefont{S.}~\bibnamefont{Joubaud}},
  \bibinfo{author}{\bibfnamefont{N.~B.} \bibnamefont{Garnier}},
  \bibnamefont{and}
  \bibinfo{author}{\bibfnamefont{S.}~\bibnamefont{Ciliberto}},
  \bibinfo{journal}{J. Stat. Mech.} \textbf{\bibinfo{volume}{2007}},
  \bibinfo{pages}{P09018} (\bibinfo{year}{2007}), ISSN
  \bibinfo{issn}{1742-5468}, \bibinfo{note}{publisher: IOP Publishing},
  \urlprefix\url{https://doi.org/10.1088/1742-5468/2007/09/p09018}.

\bibitem[{\citenamefont{Kwon et~al.}(2013)\citenamefont{Kwon, Noh, and
  Park}}]{Kwon2013}
\bibinfo{author}{\bibfnamefont{C.}~\bibnamefont{Kwon}},
  \bibinfo{author}{\bibfnamefont{J.~D.} \bibnamefont{Noh}}, \bibnamefont{and}
  \bibinfo{author}{\bibfnamefont{H.}~\bibnamefont{Park}},
  \bibinfo{journal}{Physical Review E - Statistical, Nonlinear, and Soft Matter
  Physics} \textbf{\bibinfo{volume}{88}}, \bibinfo{pages}{1}
  (\bibinfo{year}{2013}), ISSN \bibinfo{issn}{15393755}.

\bibitem[{\citenamefont{Taniguchi and Cohen}(2008)}]{taniguchi2008inertial}
\bibinfo{author}{\bibfnamefont{T.}~\bibnamefont{Taniguchi}} \bibnamefont{and}
  \bibinfo{author}{\bibfnamefont{E.}~\bibnamefont{Cohen}},
  \bibinfo{journal}{Journal of Statistical Physics}
  \textbf{\bibinfo{volume}{130}}, \bibinfo{pages}{1} (\bibinfo{year}{2008}).

\bibitem[{\citenamefont{Sabhapandit}(2012)}]{sabhapandit_heat_2012}
\bibinfo{author}{\bibfnamefont{S.}~\bibnamefont{Sabhapandit}},
  \bibinfo{journal}{Phys. Rev. E} \textbf{\bibinfo{volume}{85}},
  \bibinfo{pages}{021108} (\bibinfo{year}{2012}), ISSN
  \bibinfo{issn}{1539-3755, 1550-2376},
  \urlprefix\url{https://link.aps.org/doi/10.1103/PhysRevE.85.021108}.

\bibitem[{\citenamefont{Pal and Sabhapandit}(2014)}]{PhysRevE.90.052116}
\bibinfo{author}{\bibfnamefont{A.}~\bibnamefont{Pal}} \bibnamefont{and}
  \bibinfo{author}{\bibfnamefont{S.}~\bibnamefont{Sabhapandit}},
  \bibinfo{journal}{Phys. Rev. E} \textbf{\bibinfo{volume}{90}},
  \bibinfo{pages}{052116} (\bibinfo{year}{2014}),
  \urlprefix\url{https://link.aps.org/doi/10.1103/PhysRevE.90.052116}.

\bibitem[{\citenamefont{Munakata and Rosinberg}(2012)}]{munakata_entropy_2012}
\bibinfo{author}{\bibfnamefont{T.}~\bibnamefont{Munakata}} \bibnamefont{and}
  \bibinfo{author}{\bibfnamefont{M.~L.} \bibnamefont{Rosinberg}},
  \bibinfo{journal}{J. Stat. Mech.} \textbf{\bibinfo{volume}{2012}},
  \bibinfo{pages}{P05010} (\bibinfo{year}{2012}), ISSN
  \bibinfo{issn}{1742-5468}, \bibinfo{note}{publisher: IOP Publishing},
  \urlprefix\url{https://iopscience.iop.org/article/10.1088/1742-5468/2012/05/P05010/meta}.

\bibitem[{\citenamefont{Rosinberg et~al.}(2017)\citenamefont{Rosinberg, Tarjus,
  and Munakata}}]{rosinberg_stochastic_2017}
\bibinfo{author}{\bibfnamefont{M.~L.} \bibnamefont{Rosinberg}},
  \bibinfo{author}{\bibfnamefont{G.}~\bibnamefont{Tarjus}}, \bibnamefont{and}
  \bibinfo{author}{\bibfnamefont{T.}~\bibnamefont{Munakata}},
  \bibinfo{journal}{Phys. Rev. E} \textbf{\bibinfo{volume}{95}},
  \bibinfo{pages}{022123} (\bibinfo{year}{2017}), ISSN
  \bibinfo{issn}{2470-0045, 2470-0053},
  \urlprefix\url{https://link.aps.org/doi/10.1103/PhysRevE.95.022123}.

\bibitem[{\citenamefont{Wio}(2013)}]{wio2013path}
\bibinfo{author}{\bibfnamefont{H.~S.} \bibnamefont{Wio}},
  \emph{\bibinfo{title}{Path integrals for stochastic processes: An
  introduction}} (\bibinfo{publisher}{World Scientific}, \bibinfo{year}{2013}).

\bibitem[{\citenamefont{Cugliandolo and Lecomte}(2017)}]{cugliandolo2017rules}
\bibinfo{author}{\bibfnamefont{L.~F.} \bibnamefont{Cugliandolo}}
  \bibnamefont{and} \bibinfo{author}{\bibfnamefont{V.}~\bibnamefont{Lecomte}},
  \bibinfo{journal}{Journal of Physics A: Mathematical and Theoretical}
  \textbf{\bibinfo{volume}{50}}, \bibinfo{pages}{345001}
  (\bibinfo{year}{2017}).

\bibitem[{\citenamefont{Cugliandolo et~al.}(2019)\citenamefont{Cugliandolo,
  Lecomte, and Van~Wijland}}]{cugliandolo2019building}
\bibinfo{author}{\bibfnamefont{L.~F.} \bibnamefont{Cugliandolo}},
  \bibinfo{author}{\bibfnamefont{V.}~\bibnamefont{Lecomte}}, \bibnamefont{and}
  \bibinfo{author}{\bibfnamefont{F.}~\bibnamefont{Van~Wijland}},
  \bibinfo{journal}{Journal of Physics A: Mathematical and Theoretical}
  \textbf{\bibinfo{volume}{52}}, \bibinfo{pages}{50LT01}
  (\bibinfo{year}{2019}).

\bibitem[{\citenamefont{Kleinert}(1986)}]{kleinert_path_1986}
\bibinfo{author}{\bibfnamefont{H.}~\bibnamefont{Kleinert}},
  \bibinfo{journal}{Journal of Mathematical Physics}
  \textbf{\bibinfo{volume}{27}}, \bibinfo{pages}{3003} (\bibinfo{year}{1986}),
  ISSN \bibinfo{issn}{0022-2488, 1089-7658},
  \urlprefix\url{http://aip.scitation.org/doi/10.1063/1.527228}.

\bibitem[{\citenamefont{Chouchaoui et~al.}(1993)\citenamefont{Chouchaoui,
  Chetouani, and Hammann}}]{chouchaoui_path_1993}
\bibinfo{author}{\bibfnamefont{A.}~\bibnamefont{Chouchaoui}},
  \bibinfo{author}{\bibfnamefont{L.}~\bibnamefont{Chetouani}},
  \bibnamefont{and} \bibinfo{author}{\bibfnamefont{T.~F.}
  \bibnamefont{Hammann}}, \bibinfo{journal}{Fortschr. Phys.}
  \textbf{\bibinfo{volume}{41}}, \bibinfo{pages}{201} (\bibinfo{year}{1993}),
  ISSN \bibinfo{issn}{00158209, 15213979},
  \urlprefix\url{http://doi.wiley.com/10.1002/prop.2190410303}.

\bibitem[{\citenamefont{Suassuna et~al.}(2021)\citenamefont{Suassuna, Melo, and
  Guerreiro}}]{suassuna_path_2021}
\bibinfo{author}{\bibfnamefont{B.}~\bibnamefont{Suassuna}},
  \bibinfo{author}{\bibfnamefont{B.}~\bibnamefont{Melo}}, \bibnamefont{and}
  \bibinfo{author}{\bibfnamefont{T.}~\bibnamefont{Guerreiro}},
  \bibinfo{journal}{Phys. Rev. A} \textbf{\bibinfo{volume}{103}},
  \bibinfo{pages}{013110} (\bibinfo{year}{2021}), \bibinfo{note}{publisher:
  American Physical Society},
  \urlprefix\url{https://link.aps.org/doi/10.1103/PhysRevA.103.013110}.

\bibitem[{\citenamefont{Chaichian and Demichev}(2018)}]{chaichian2018path}
\bibinfo{author}{\bibfnamefont{M.}~\bibnamefont{Chaichian}} \bibnamefont{and}
  \bibinfo{author}{\bibfnamefont{A.}~\bibnamefont{Demichev}},
  \emph{\bibinfo{title}{Path integrals in physics: Volume I stochastic
  processes and quantum mechanics}} (\bibinfo{publisher}{CRC Press},
  \bibinfo{year}{2018}).

\bibitem[{\citenamefont{Douarche et~al.}(2006)\citenamefont{Douarche, Joubaud,
  Garnier, Petrosyan, and Ciliberto}}]{douarche_work_2006}
\bibinfo{author}{\bibfnamefont{F.}~\bibnamefont{Douarche}},
  \bibinfo{author}{\bibfnamefont{S.}~\bibnamefont{Joubaud}},
  \bibinfo{author}{\bibfnamefont{N.~B.} \bibnamefont{Garnier}},
  \bibinfo{author}{\bibfnamefont{A.}~\bibnamefont{Petrosyan}},
  \bibnamefont{and}
  \bibinfo{author}{\bibfnamefont{S.}~\bibnamefont{Ciliberto}},
  \bibinfo{journal}{Phys. Rev. Lett.} \textbf{\bibinfo{volume}{97}},
  \bibinfo{pages}{140603} (\bibinfo{year}{2006}), \bibinfo{note}{publisher:
  American Physical Society},
  \urlprefix\url{https://link.aps.org/doi/10.1103/PhysRevLett.97.140603}.

\bibitem[{\citenamefont{Mart{\'\i}nez et~al.}(2016)\citenamefont{Mart{\'\i}nez,
  Rold{\'a}n, Dinis, Petrov, Parrondo, and Rica}}]{martinez2016brownian}
\bibinfo{author}{\bibfnamefont{I.~A.} \bibnamefont{Mart{\'\i}nez}},
  \bibinfo{author}{\bibfnamefont{{\'E}.}~\bibnamefont{Rold{\'a}n}},
  \bibinfo{author}{\bibfnamefont{L.}~\bibnamefont{Dinis}},
  \bibinfo{author}{\bibfnamefont{D.}~\bibnamefont{Petrov}},
  \bibinfo{author}{\bibfnamefont{J.~M.} \bibnamefont{Parrondo}},
  \bibnamefont{and} \bibinfo{author}{\bibfnamefont{R.~A.} \bibnamefont{Rica}},
  \bibinfo{journal}{Nature physics} \textbf{\bibinfo{volume}{12}},
  \bibinfo{pages}{67} (\bibinfo{year}{2016}).

\bibitem[{\citenamefont{Holubec and Ryabov}(2021)}]{holubec2021fluctuations}
\bibinfo{author}{\bibfnamefont{V.}~\bibnamefont{Holubec}} \bibnamefont{and}
  \bibinfo{author}{\bibfnamefont{A.}~\bibnamefont{Ryabov}},
  \bibinfo{journal}{Journal of Physics A: Mathematical and Theoretical}
  \textbf{\bibinfo{volume}{55}}, \bibinfo{pages}{013001}
  (\bibinfo{year}{2021}).

\bibitem[{\citenamefont{Lemons and Gythiel}(1997)}]{lemons1997paul}
\bibinfo{author}{\bibfnamefont{D.~S.} \bibnamefont{Lemons}} \bibnamefont{and}
  \bibinfo{author}{\bibfnamefont{A.}~\bibnamefont{Gythiel}},
  \bibinfo{journal}{American Journal of Physics} \textbf{\bibinfo{volume}{65}},
  \bibinfo{pages}{1079} (\bibinfo{year}{1997}).

\bibitem[{\citenamefont{Sekimoto}(1998)}]{sekimoto1998langevin}
\bibinfo{author}{\bibfnamefont{K.}~\bibnamefont{Sekimoto}},
  \bibinfo{journal}{Progress of Theoretical Physics Supplement}
  \textbf{\bibinfo{volume}{130}}, \bibinfo{pages}{17} (\bibinfo{year}{1998}).

\bibitem[{\citenamefont{Singh}(2008)}]{singh_onsager-machlup_2008}
\bibinfo{author}{\bibfnamefont{N.}~\bibnamefont{Singh}}, \bibinfo{journal}{J
  Stat Phys} \textbf{\bibinfo{volume}{131}}, \bibinfo{pages}{405}
  (\bibinfo{year}{2008}), ISSN \bibinfo{issn}{1572-9613},
  \urlprefix\url{https://doi.org/10.1007/s10955-008-9503-5}.

\bibitem[{\citenamefont{Wampler and Barato}(2021)}]{wampler2021skewness}
\bibinfo{author}{\bibfnamefont{T.}~\bibnamefont{Wampler}} \bibnamefont{and}
  \bibinfo{author}{\bibfnamefont{A.~C.} \bibnamefont{Barato}},
  \bibinfo{journal}{Journal of Physics A: Mathematical and Theoretical}
  \textbf{\bibinfo{volume}{55}}, \bibinfo{pages}{014002}
  (\bibinfo{year}{2021}).

\bibitem[{\citenamefont{Arold et~al.}(2018)\citenamefont{Arold, Dechant, and
  Lutz}}]{arold_heat_2018}
\bibinfo{author}{\bibfnamefont{D.}~\bibnamefont{Arold}},
  \bibinfo{author}{\bibfnamefont{A.}~\bibnamefont{Dechant}}, \bibnamefont{and}
  \bibinfo{author}{\bibfnamefont{E.}~\bibnamefont{Lutz}},
  \bibinfo{journal}{Phys. Rev. E} \textbf{\bibinfo{volume}{97}},
  \bibinfo{pages}{022131} (\bibinfo{year}{2018}), \bibinfo{note}{publisher:
  American Physical Society},
  \urlprefix\url{https://link.aps.org/doi/10.1103/PhysRevE.97.022131}.

\bibitem[{\citenamefont{Celani et~al.}(2012)\citenamefont{Celani, Bo, Eichhorn,
  and Aurell}}]{celani_anomalous_2012}
\bibinfo{author}{\bibfnamefont{A.}~\bibnamefont{Celani}},
  \bibinfo{author}{\bibfnamefont{S.}~\bibnamefont{Bo}},
  \bibinfo{author}{\bibfnamefont{R.}~\bibnamefont{Eichhorn}}, \bibnamefont{and}
  \bibinfo{author}{\bibfnamefont{E.}~\bibnamefont{Aurell}},
  \bibinfo{journal}{Phys. Rev. Lett.} \textbf{\bibinfo{volume}{109}},
  \bibinfo{pages}{260603} (\bibinfo{year}{2012}), ISSN
  \bibinfo{issn}{0031-9007, 1079-7114},
  \urlprefix\url{https://link.aps.org/doi/10.1103/PhysRevLett.109.260603}.

\bibitem[{\citenamefont{Dechant et~al.}(2017)\citenamefont{Dechant, Kiesel, and
  Lutz}}]{dechant_underdamped_2017}
\bibinfo{author}{\bibfnamefont{A.}~\bibnamefont{Dechant}},
  \bibinfo{author}{\bibfnamefont{N.}~\bibnamefont{Kiesel}}, \bibnamefont{and}
  \bibinfo{author}{\bibfnamefont{E.}~\bibnamefont{Lutz}},
  \bibinfo{journal}{EPL} \textbf{\bibinfo{volume}{119}}, \bibinfo{pages}{50003}
  (\bibinfo{year}{2017}), ISSN \bibinfo{issn}{0295-5075},
  \bibinfo{note}{publisher: IOP Publishing},
  \urlprefix\url{https://doi.org/10.1209/0295-5075/119/50003}.

\bibitem[{\citenamefont{Bo and Celani}(2013)}]{bo2013entropic}
\bibinfo{author}{\bibfnamefont{S.}~\bibnamefont{Bo}} \bibnamefont{and}
  \bibinfo{author}{\bibfnamefont{A.}~\bibnamefont{Celani}},
  \bibinfo{journal}{Physical Review E} \textbf{\bibinfo{volume}{87}},
  \bibinfo{pages}{050102} (\bibinfo{year}{2013}).

\bibitem[{\citenamefont{Schmiedl and Seifert}(2007)}]{schmiedl2007efficiency}
\bibinfo{author}{\bibfnamefont{T.}~\bibnamefont{Schmiedl}} \bibnamefont{and}
  \bibinfo{author}{\bibfnamefont{U.}~\bibnamefont{Seifert}},
  \bibinfo{journal}{EPL (Europhysics Letters)} \textbf{\bibinfo{volume}{81}},
  \bibinfo{pages}{20003} (\bibinfo{year}{2007}).

\bibitem[{\citenamefont{Plata et~al.}(2020)\citenamefont{Plata,
  Gu{\'e}ry-Odelin, Trizac, and Prados}}]{plata2020building}
\bibinfo{author}{\bibfnamefont{C.~A.} \bibnamefont{Plata}},
  \bibinfo{author}{\bibfnamefont{D.}~\bibnamefont{Gu{\'e}ry-Odelin}},
  \bibinfo{author}{\bibfnamefont{E.}~\bibnamefont{Trizac}}, \bibnamefont{and}
  \bibinfo{author}{\bibfnamefont{A.}~\bibnamefont{Prados}},
  \bibinfo{journal}{Journal of Statistical Mechanics: Theory and Experiment}
  \textbf{\bibinfo{volume}{2020}}, \bibinfo{pages}{093207}
  (\bibinfo{year}{2020}).

\bibitem[{\citenamefont{Fogedby}(2020{\natexlab{b}})}]{Fogedby_2020}
\bibinfo{author}{\bibfnamefont{H.~C.} \bibnamefont{Fogedby}},
  \bibinfo{journal}{Journal of Statistical Mechanics: Theory and Experiment}
  \textbf{\bibinfo{volume}{2020}}, \bibinfo{pages}{083208}
  (\bibinfo{year}{2020}{\natexlab{b}}),
  \urlprefix\url{https://doi.org/10.1088/1742-5468/aba7b2}.

\bibitem[{\citenamefont{Coffey and Kalmykov}(2012)}]{coffey2012langevin}
\bibinfo{author}{\bibfnamefont{W.}~\bibnamefont{Coffey}} \bibnamefont{and}
  \bibinfo{author}{\bibfnamefont{Y.~P.} \bibnamefont{Kalmykov}},
  \emph{\bibinfo{title}{The Langevin equation: with applications to stochastic
  problems in physics, chemistry and electrical engineering}},
  vol.~\bibinfo{volume}{27} (\bibinfo{publisher}{World Scientific},
  \bibinfo{year}{2012}).

\bibitem[{\citenamefont{Mori}(1965)}]{Mori1965}
\bibinfo{author}{\bibfnamefont{H.}~\bibnamefont{Mori}}, \bibinfo{journal}{Prog.
  Theor. Phys.} \textbf{\bibinfo{volume}{33}}, \bibinfo{pages}{423}
  (\bibinfo{year}{1965}).

\bibitem[{\citenamefont{Ro{\ss}nagel et~al.}(2016)\citenamefont{Ro{\ss}nagel,
  Dawkins, Tolazzi, Abah, Lutz, Schmidt-Kaler, and Singer}}]{Rossnagel2016}
\bibinfo{author}{\bibfnamefont{J.}~\bibnamefont{Ro{\ss}nagel}},
  \bibinfo{author}{\bibfnamefont{S.~T.} \bibnamefont{Dawkins}},
  \bibinfo{author}{\bibfnamefont{K.~N.} \bibnamefont{Tolazzi}},
  \bibinfo{author}{\bibfnamefont{O.}~\bibnamefont{Abah}},
  \bibinfo{author}{\bibfnamefont{E.}~\bibnamefont{Lutz}},
  \bibinfo{author}{\bibfnamefont{F.}~\bibnamefont{Schmidt-Kaler}},
  \bibnamefont{and} \bibinfo{author}{\bibfnamefont{K.}~\bibnamefont{Singer}},
  \bibinfo{journal}{Science} \textbf{\bibinfo{volume}{352}},
  \bibinfo{pages}{325} (\bibinfo{year}{2016}), ISSN \bibinfo{issn}{10959203},
  \eprint{1510.03681}.

\bibitem[{\citenamefont{Nascimento and Morgado}(2019)}]{NascimentoMorgado2019}
\bibinfo{author}{\bibfnamefont{E.~S.} \bibnamefont{Nascimento}}
  \bibnamefont{and} \bibinfo{author}{\bibfnamefont{W.~A.~M.}
  \bibnamefont{Morgado}}, \bibinfo{journal}{{EPL} (Europhysics Letters)}
  \textbf{\bibinfo{volume}{126}}, \bibinfo{pages}{10002}
  (\bibinfo{year}{2019}).

\bibitem[{\citenamefont{Mathai}(1993)}]{mathai1993handbook}
\bibinfo{author}{\bibfnamefont{A.~M.} \bibnamefont{Mathai}},
  \emph{\bibinfo{title}{A handbook of generalized special functions for
  statistical and physical sciences}} (\bibinfo{publisher}{Oxford University
  Press, USA}, \bibinfo{year}{1993}).

\bibitem[{\citenamefont{Blickle and Bechinger}(2012)}]{blickle2012realization}
\bibinfo{author}{\bibfnamefont{V.}~\bibnamefont{Blickle}} \bibnamefont{and}
  \bibinfo{author}{\bibfnamefont{C.}~\bibnamefont{Bechinger}},
  \bibinfo{journal}{Nature Physics} \textbf{\bibinfo{volume}{8}},
  \bibinfo{pages}{143} (\bibinfo{year}{2012}).

\bibitem[{\citenamefont{Onsager and Machlup}(1953)}]{onsager1953fluctuations}
\bibinfo{author}{\bibfnamefont{L.}~\bibnamefont{Onsager}} \bibnamefont{and}
  \bibinfo{author}{\bibfnamefont{S.}~\bibnamefont{Machlup}},
  \bibinfo{journal}{Physical Review} \textbf{\bibinfo{volume}{91}},
  \bibinfo{pages}{1505} (\bibinfo{year}{1953}).

\bibitem[{\citenamefont{Bo et~al.}(2019)\citenamefont{Bo, Lim, and
  Eichhorn}}]{bo2019functionals}
\bibinfo{author}{\bibfnamefont{S.}~\bibnamefont{Bo}},
  \bibinfo{author}{\bibfnamefont{S.~H.} \bibnamefont{Lim}}, \bibnamefont{and}
  \bibinfo{author}{\bibfnamefont{R.}~\bibnamefont{Eichhorn}},
  \bibinfo{journal}{Journal of Statistical Mechanics: Theory and Experiment}
  \textbf{\bibinfo{volume}{2019}}, \bibinfo{pages}{084005}
  (\bibinfo{year}{2019}).

\bibitem[{\citenamefont{Moreno et~al.}(2019)\citenamefont{Moreno, Barci, and
  Arenas}}]{moreno2019conditional}
\bibinfo{author}{\bibfnamefont{M.~V.} \bibnamefont{Moreno}},
  \bibinfo{author}{\bibfnamefont{D.~G.} \bibnamefont{Barci}}, \bibnamefont{and}
  \bibinfo{author}{\bibfnamefont{Z.~G.} \bibnamefont{Arenas}},
  \bibinfo{journal}{Physical Review E} \textbf{\bibinfo{volume}{99}},
  \bibinfo{pages}{032125} (\bibinfo{year}{2019}).

\bibitem[{\citenamefont{Feynman et~al.}(2010)\citenamefont{Feynman, Hibbs, and
  Styer}}]{feynman2010quantum}
\bibinfo{author}{\bibfnamefont{R.~P.} \bibnamefont{Feynman}},
  \bibinfo{author}{\bibfnamefont{A.~R.} \bibnamefont{Hibbs}}, \bibnamefont{and}
  \bibinfo{author}{\bibfnamefont{D.~F.} \bibnamefont{Styer}},
  \emph{\bibinfo{title}{Quantum mechanics and path integrals}}
  (\bibinfo{publisher}{Courier Corporation}, \bibinfo{year}{2010}).

\bibitem[{\citenamefont{Bender et~al.}(1999)\citenamefont{Bender, Orszag, and
  Orszag}}]{bender1999advanced}
\bibinfo{author}{\bibfnamefont{C.}~\bibnamefont{Bender}},
  \bibinfo{author}{\bibfnamefont{S.}~\bibnamefont{Orszag}}, \bibnamefont{and}
  \bibinfo{author}{\bibfnamefont{S.}~\bibnamefont{Orszag}},
  \emph{\bibinfo{title}{Advanced Mathematical Methods for Scientists and
  Engineers I: Asymptotic Methods and Perturbation Theory}}, Advanced
  Mathematical Methods for Scientists and Engineers
  (\bibinfo{publisher}{Springer}, \bibinfo{year}{1999}), ISBN
  \bibinfo{isbn}{9780387989310},
  \urlprefix\url{https://books.google.com.br/books?id=-yQXwhE6iWMC}.

\end{thebibliography}

\end{document}